\documentclass[%
reprint,
superscriptaddress,
nofootinbib,
 amsmath,amssymb,
 aps,
floatfix,
prx
]{revtex4-2}

\usepackage{graphicx}% Include figure files
\usepackage{dcolumn}% Align table columns on decimal point
\usepackage{bm}% bold math
\usepackage{cancel}
\usepackage[separate-uncertainty=true,range-phrase=--,multi-part-units=single,range-units=single]{siunitx}
\newcommand{\apx}{Appendix}
\newcommand{\apxs}{Appendices}
\newcommand{\avg}[1]{\left\langle#1\right\rangle}
\newcommand{\psigma}{p_{\sigma}}
\newcommand{\pcalt}{p_{\mathcal{T}}}
\newcommand{\kB}{k_{\text{B}}}
\newcommand{\T}{\mathrm{T}}

\DeclareMathOperator{\const}{const}
\DeclareMathOperator{\Var}{Var}
\usepackage{hyperref}% add hypertext capabilities
\hypersetup{
    colorlinks,
    linkcolor={blue},
    citecolor={blue},
    urlcolor={blue}
}
\usepackage{orcidlink}
\makeatletter
\renewcommand{\paragraph}{\@startsection{paragraph}{4}{\z@}%
  {-3.25ex \@plus -1ex \@minus -0.2ex}%
  {0ex}%
  {\normalfont\normalsize\itshape}}
\makeatother
\begin{document}

\preprint{APS/123-QED}

% \title{Manuscript Title:\\with Forced Linebreak}% Force line breaks with \\
%\title{Microcanonical principle of maximum caliber for spatially extended, nonequilibrium, and active thermodynamic systems}% Force line breaks with \\
%\title{Microcanonical principle of maximum caliber}
%\title{Maximum caliber in microcanonical ensemble out of equilibrium}
\title{Microcanonical ensemble out of equilibrium}

% \author{Ann Author}
%  \altaffiliation[Also at ]{Physics Department, XYZ University.}%Lines break automatically or can be forced with \\
%  \email{Second.Author@institution.edu}
% \affiliation{%
%  Authors' institution and/or address\\
%  This line break forced with \textbackslash\textbackslash
% }%

% \collaboration{MUSO Collaboration}%\noaffiliation

%  \homepage{http://www.Second.institution.edu/~Charlie.Author}
% \affiliation{
%  Second institution and/or address\\
%  This line break forced% with \\
% }%
% \affiliation{
%  Third institution, the second for Charlie Author
% \affiliation{%
%  Authors' institution and/or address\\
%  This line break forced with \textbackslash\textbackslash
% }%

% \collaboration{CLEO Collaboration}%\noaffiliation

\author{R. Belousov\orcidlink{0000-0002-8896-8109}}
\email{roman.belousov@embl.de}
\affiliation{Cell Biology and Biophysics Unit, European Molecular Biology Laboratory, Meyerhofstraße 1, 69117 Heidelberg, Germany}
\author{J. Elliott\orcidlink{0000-0002-3703-4234}}
\affiliation{Cell Biology and Biophysics Unit, European Molecular Biology Laboratory, Meyerhofstraße 1, 69117 Heidelberg, Germany}
\affiliation{Department of Physics and Astronomy, Heidelberg University, 69120 Heidelberg, Germany}
\author{F. Berger\orcidlink{0000-0003-3355-4336}}
\affiliation{Cell Biology, Neurobiology and Biophysics, Department of Biology, Faculty of Science, Utrecht University, 3584 Utrecht, The Netherlands}
\author{L. Rondoni\orcidlink{0000-0002-4223-6279}}
\affiliation{Department of Mathematical Sciences, Politecnico di Torino, Corso Duca degli Abruzzi 24, 10129 Turin, Italy}
\affiliation{INFN, Sezione di Torino, Turin 10125, Italy}
\author{A. Erzberger\orcidlink{0000-0002-2200-4665}}%
\email{erzberge@embl.de}
\affiliation{Cell Biology and Biophysics Unit, European Molecular Biology Laboratory, Meyerhofstraße 1, 69117 Heidelberg, Germany}
\affiliation{Department of Physics and Astronomy, Heidelberg University, 69120 Heidelberg, Germany}

\date{\today}% It is always \today, today,

\begin{abstract}
Introduced by Boltzmann under the name ``monode,'' the microcanonical ensemble serves as the fundamental representation of equilibrium thermodynamics in statistical mechanics by counting all possible realizations of a system's states. Ensemble theory connects this idea with probability and information theory, leading to the notion of Shannon-Gibbs entropy and, ultimately, to the principle of maximum caliber describing \textit{trajectories} of systems---in and out of equilibrium. While the latter phenomenological generalization reproduces many results of nonequilibrium thermodynamics, given a proper choice of observables, its physical justification remains an open area of research. What is the microscopic origin and physical interpretation of this variational approach? What guides the choice of relevant observables? We address these questions by extending Boltzmann's method to a microcanonical caliber principle and counting realizations of a system's trajectories---all assumed equally probable. Maximizing the microcanonical caliber under the imposed constraints, we systematically develop generalized detailed-balance relations, clarify the statistical origins of inhomogeneous transport, and provide an independent derivation of key equations from stochastic thermodynamics. This approach introduces a \textit{dynamical} ensemble theory for nonequilibrium steady states in spatially extended and active systems. While verifying the equivalence of ensembles, e.g. those of Norton and Th\'evenin, our framework contests other common assumptions about nonequilibrium regimes, with supporting evidence provided by stochastic simulations. Our theory suggests further connections to the first principles of microscopic dynamics in classical statistical mechanics, which are essential for investigating systems where the necessary conditions for thermodynamic behavior are not satisfied.
\end{abstract}

\maketitle

% \tableofcontents
% \section{Introduction}
Tools of statistical mechanics have become indispensable in addressing modern problems of science and technology, which concern complex, active, and, in general, far from equilibrium systems~\cite{Kriebisch2025,Genthon2025,Haugerud2024,Wang2024,Bauermann2022,Bauermann2023,Liu2023,Zakine2023,Bowick2022,Markovich2021,Gaspard2020,Herpich2018,Sartori2019,Benoist2025,Takatori2015,Chetrite2014,Freitas2021,Petridou2021,Tang2020,Tang2021,Li2023,Avanzini2024,Zheng2024,Raghu2025,Tkaik2025}. These tools extend, typically in a phenomenological manner, classical results based on the theory of equilibrium thermodynamic systems, e.g.\ local detailed-balance relations~\cite{Maes2020,Maes2021,Evans2005}, statistics of large deviations~\cite{Falasco2025,Touchette2018,Touchette2013,Smith2011,Touchette2009}, response theory~\cite{Gaspard2013,Ruelle2009,Schnakenberg1976}. The assumptions implied in such extensions are difficult to validate and, when not verified, drastically limit applications or physical interpretation of theoretical models. To overcome the limitations of phenomenological approaches to nonequilibrium thermodynamics, the search for generalizations of the principles, which statistical mechanics established for equilibrium systems---ensemble methods~\cite{Ghosh2020,Pradhan2010}, Helmholtz' mechanical interpretation~\cite{Porporato2024,Cardin2004}, and dynamical-systems theory~\cite{Ding2025,Caruso2020,Dhar2019,Agarwal2015,Ruelle2011,Lebowitz1957,Bergmann1955},---continues.

The ensemble theory of statistical mechanics~\cite{Gibbs1902,Callen1991} provides the most widely adopted framework of thermodynamics, which relates macroscopic properties of \textit{equilibrium systems} to time-independent statistics of their micro- or mesoscopic states $\bm{s} \in \bm{\sigma}$, e.g.\ positions and momenta of all the particles or occupation numbers of energy levels. Dynamical fluctuations in these systems are thus characterized by the probability distribution $\psigma(\bm{s})$, which does not change in time, because any macroscopic variable $\avg{A(\bm{s})} = \int_{\bm{\sigma}} \bm{ds}\, \psigma(\bm{s}) A(\bm{s})$ must by definition remain constant in equilibrium. According to statistical mechanics, the distribution~$\psigma(\bm{s})$ maximizes the Gibbs entropy defined as
\begin{equation}\label{eq:shannon}
    S_{\sigma} = - \kB \int_{\bm{\sigma}} \bm{ds}\, \psigma(\bm{s}) \ln \psigma(\bm{s}),
\end{equation}
where the Boltzmann constant, $\kB=\SI{1.380649e-23}{J/K}$, has units of entropy and quantifies the vast disparity of scales between the macroscopic and the microscopic realms. In systems with discrete sets of states, the integral is replaced by the summation.

If one does not speak of the thermodynamic entropy, which is an objective function of state, $S_\sigma / \kB$ can be interpreted as a measure of indeterminacy. In this context, it is commonly referred to as Shannon entropy, a subjective quantity that depends on the level of system description. For instance, a level of uncertainty can be intentionally introduced through coarse graining~\cite{Esposito2012,Puglisi2010}, as suitable in some applications~\cite{Harunari2022,Altaner2012}.

As an extension of the information-theoretic approach from time-independent macroscopic states to general dynamical processes, the principle of maximum caliber was coined by \citet{Jaynes1980}, although its foundations go back to much earlier works of his~\cite{Jaynes1957}, as well as of \citet{Filyukov1967I,Filyukov1967II}. If we assign probabilities $\pcalt(\bm{s}_t)$ to the possible trajectories $\bm{s}_t$---a time-ordered succession of states---of a system, its caliber or \textit{path} entropy over the interval $t \in \mathcal{T}$ mimics Eq.~\eqref{eq:shannon} as
\begin{equation}\label{eq:jaynes}
    S_{\mathcal{T}} = -\int_{\bm{\sigma} \times \mathcal{T}} \mathcal{D}\bm{s}_t\, \pcalt(\bm{s}_t) \ln \pcalt(\bm{s}_t),
\end{equation}
with the integral taken over all possible paths $s_t$. According to the principle of maximum caliber or path entropy,\footnote{or second entropy~\cite{Attard2009,Attard2012}} which is thought to be applicable also out of equilibrium~\cite{Filyukov1967I,Filyukov1967II,Filyukov1968,Jaynes1980,Haken1986II,Dewar2003,Gaspard2004,Evans2005,Dewar2005,Ghosh2006,Lecomte2007,Seitaridou2007,Attard2009,Wu2009,Otten2010,Stock2008,Press2010,Monthus2011,Ge2012,Press2011,Smith2011,Ghosh2011,Lee2012I,Lee2012II,Press2013,Hazoglou2015,Ghosh2020}, the probability distribution $\pcalt(\bm{s}_t)$ is identified by extremizing Eq.~\eqref{eq:shannon} subject to known macroscopic constraints (\apx~\ref{app:variation}).

Despite the mathematical analogy shared by Eqs.~\eqref{eq:shannon} and \eqref{eq:jaynes}, the principle of maximum caliber still lacks the foundations that are well established and understood within the context of maximum-entropy principle for equilibrium systems, whose relevant variables and constraints are known \cite{Auletta2017}. In particular, a complete picture of nonequilibrium statistics---the mathematical derivation and its physical interpretation based on the dynamics of microscopic constituents of matter~\cite{CRV2014}---is so far missing. Statistical mechanics encompasses the relevant macroscopic constraints that the equilibrium ensembles must verify, consistently with Liouville's theorem and Hamiltonian dynamics~\cite[Chapter 2]{Tuckerman2023}. Consequently, statistics of equilibrium systems can be derived from the microscopic representation of these constraints within the microcanonical ensemble~\cite[Chapter 3]{Tuckerman2023}, \cite{Pearson1985}.

When maximizing the system's caliber, the choice of the macroscopic constraints is typically guided by phenomenological laws of nonequilibrium thermodynamics. Thereby the selected variables often include---in addition to properties whose relevance is well established from the equilibrium theory, like energy and mass,---phenomenological parameters such as the magnitudes of currents. However the distinction between these types of variables is not always appreciated and may lead to an inadequate, or valid only in a certain limit, description.

Even for nonequilibrium steady states---regimes formally similar to  equilibrium, because they are characterized by constant nonzero fluxes---there is no universally accepted choice of the macroscopic statistics. Variants of the maximum-caliber formalism differ by imposed constraints or by the way interactions with thermodynamic reservoirs are modeled~\cite{Filyukov1967II,Evans2004I,Evans2004II,Stock2008,Press2010,Monthus2011,Ge2012,Pachter2023}. In fact, as we find here, even the commonly invoked condition of a constant macroscopic flux can arise from different physical mechanisms, leading to distinct nonequilibrium regimes.

Especially when multiple constraints must be chosen to describe a complex system, physical insights are in high demand. For example, which macroscopic observables would provide the relevant statistics for nontrivial spatial organization---such as gradients of chemical components or patterns in living systems,---that shapes important processes in biology and engineering~\cite{Rombouts2023,Davies2020,Hua2021,Halatek2018,Keil2017,Manning1961}? This question leads us to consider also applications of the maximum-caliber principle to systems whose thermodynamic variables can be regarded as fields. In particular, we show how the symmetry broken by nonequilibrium mechanisms entails a spatial gradient of matter.

To develop the ensemble representation for systems beyond the scope of classical equilibrium thermodynamics, we combine \textit{some} ideas of the caliber theory~\cite{Ghosh2006} with the microcanonical approach, starting from Boltzmann's simplified model of an ideal gas~\cite{Sharp2015}. We proceed by counting all possible microscopic trajectories of a system---assumed equally likely---rather than its microscopic states as in Boltzmann's original method. The macroscopic evolution of the system is then identified with its most probable path that maximizes the number of such realizations.

Subsequently, we extend our theory to more complex systems in and out of equilibrium, by incorporating more realistic dynamical constraints, which introduce interactions between the microscopic constituents of matter, ensure the continuity of its flow, or apply driving forces. In the outcome we systematically obtain by direct calculations a physically interpretable, mechanistic representation of thermodynamic ensembles for spatially extended, nonequilibrium, and active systems, which we ultimately verify in stochastic simulations.

The connection to microscopic events offers the physical insight, which is missing in the formal variational principles, and a systemetic way to complete statistical-mechanics models with conjugate pairs of thermodynamic variables suitable to describe specific nonequilibrium regimes. These variables uniquely identify the statistical ensemble, which turns out to depend not only on the magnitude and direction of the currents, e.g.\ as assumed in the typical phenomenological theories, but on the details of the driving mechanism.

The distinct features of our theory comprise the microcanonical constraints and \textit{local}, in time, analysis of trajectories. First, we impose conservation of the total energy $E$ and the number of particles $N$ in the system---regardless of any nonequilibrium driving forces $\bm\Theta$ that may be present. Second, we count the realizations of the trajectories over a short, or rather infinitesimal, time step of size $dt$. Under these conditions we replace Eqs.~\eqref{eq:shannon} and \eqref{eq:jaynes} with their microcanonical counterparts---the Boltzmann entropy $S_{\rm B} = \kB \ln W(N, E)$ and \textit{microcanonical caliber}
\begin{equation}
    S_{dt} = \ln \Omega(E, N, \bm\Theta...)
\end{equation}
where $W$ and $\Omega$ count the realizations of the system's macroscopic states and trajectories respectively.

Maximization of the microcanonical caliber subtly differs from the nonlocal variational problem formulated for the traditional path entropy over the time interval $\mathcal{T}$. The nonlocal approach identifies the most likely evolution of a system by extremizing the functional~Eq.~\eqref{eq:jaynes}, which in continuum problems typically entails the conditions for extrema at each instant of time as well, cf.\ the Euler-Lagrange equations minimizing the action functional in the classical mechanics. Therefore our microcanonical formalism is compatible with the traditional principle of maximum caliber in the macroscopic continuum limit.

%%%
\section{\label{sec:framework}Paradigmatic models}
First we develop and demonstrate our theoretical framework using two paradigmatic models of classical statistical mechanics. We begin with the Boltzmann gas, representing the simplest class of problems considered, and then proceed to the Fermi gas, which illustrates nontrivial effects of microscopic interactions.

\subsection{\label{sec:ideal}Boltzmann gas}
\begin{figure}[!t]\centering
\includegraphics[width=\columnwidth]{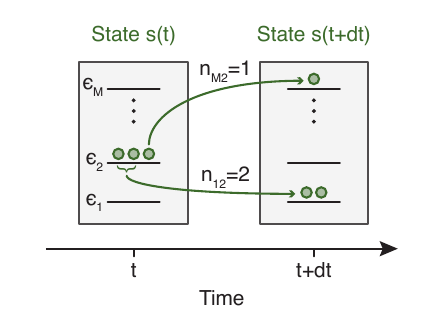}% Here is how to import EPS art
\caption{\label{fig:ideal}Simplified model of ideal gas. In the initial state $\bm{s}(t)$ indistinguishable particles are distributed over $M$ energy levels $\epsilon_{i=1,2,...,M}$. In each discrete time step $dt$, $n_{ji}$ particles move from the $i$\textsuperscript{th} to the $j$\textsuperscript{th} level, with $n_{ii}$ particles remaining in their place.}
\end{figure}

To introduce the microcanonical caliber theory in the simplest form, we begin with Boltzmann's paradigmatic model of an ideal gas in discrete time $t$ and state space $\bm{\sigma}$ (Fig~\ref{fig:ideal}): $N$ indistinguishable particles may occupy any of $M \gg N$ energy levels $\epsilon_{i=1,2,...,M}$ with degeneracies $g_i$. The system's state can thus be described by a vector of occupation numbers $\bm{s} = (n_1, n_2, ..., n_M)^{\T} \in \bm{\sigma}$.

Starting with an arbitrary initial state---not necessarily fulfilling the maximum-entropy condition (\apx~\ref{app:boltzmann})---and implementing the microcanonical conditions, we impose a \textit{macroscopic constraint} on the total energy, $E = \sum_i \epsilon_i n_i$, which must be conserved in time, and a single \textit{microscopic constraint} on the dynamics---particles are neither
created nor annihilated. They are however allowed to move to, or remain still in, any energy level, as long as in a time step $t \to t + dt$ the transition between two states $\bm{s}, \bm{s}' \in \bm{\sigma}$,  obeys $N(\bm{s}) = N(\bm{s}')$ and $E(\bm{s}) = E(\bm{s}')$. How many distinct realizations of such a transition are possible?

Given the above conditions, any transition $\bm{s}_{dt}: \bm{s} \to \bm{s}'$---or the system's \textit{path}---can be uniquely specified by the set $\{n_{ji}\}_{i,j=1}^{M}$ of numbers of particles, which jump from the $i$\textsuperscript{th} level into the $j$\textsuperscript{th} level. The elements $n_{ji}$ must satisfy the following relations:
\begin{align}\label{eq:num}
    n_i(t) -& \sum_{j=1}^M n_{ji} =0 \, ,\quad
    n_j(t + dt) - \sum_{i=1}^M n_{ji} = 0 \, ,
    \\
    \label{eq:eng}
    0 =& \sum_{ij} (\epsilon_i - \epsilon_j) n_{ji} =
    E - \sum_{j} \epsilon_j n_j(t + dt),
\end{align}
which enforce the conservation of the total number of particles and their total energy, counting also those that remain at the same level, $i = j$.

Each of the elements $n_{ji}$---which together define a path of the system---represents the number of particles taken from the state $i$ and redistributed independently over $g_j$ sublevels of energy $\epsilon_j$. The total count of indistinguishable particles' rearrangements is approximately given by the classical combinatorial formula of the Maxwell-Boltzmann statistics\footnote{The Maxwell-Boltzmann combinatorial formula may be regarded as an approximation of the Bose-Einstein statistics when $M \gg N$ and differs from the Boltzmann's exact counting of distinguishable particles by the factor of $n_i!$~\cite[Chapter 13]{Carter2001}.} $\Omega_{ji} \approx g_j^{n_{ji}}/n_{ji}!$~\cite[Chapter 13]{Carter2001}. As the particles are distributed independently, the number of realizations of the system's path $\{n_{ji}\}_{i,j=1}^{M}$ amounts to
\begin{equation}\label{eq:mb}
    \Omega_{\rm MB}\left(\bm{s}_{dt}\right) = \prod_{i,j=1}^{M} \Omega_{ji} \approx \prod_{i,j=1}^M\frac{g_j^{n_{ji}}}{n_{ji}!}.
\end{equation}

Assuming that all rearrangements of particles are equally likely, the most probable path of the system is characterized by the largest number of realizations, which is an objective quantity. It is given by the $n_{ji}$ that maximize the logarithm of Eq.~\eqref{eq:mb}---\textit{the microcanonical caliber of the path}---together with the explicit constraints on the particles' number and their energy. 
The corresponding objective function we extremize is:
\begin{multline}\label{eq:fmb}
    f_{\rm MB} = \ln \Omega_{\rm MB} + \beta\left(E - \sum_{ij} \epsilon_j n_{ji}\right) \\ + \sum_i\theta_i \left(n_i - \sum_{j} n_{ji} \right),
\end{multline}
where the Lagrange multipliers $\beta$ and $\theta_{i=1,2,...M}$ constrain the total energy $E$ and the initial occupation numbers $n_i(t)$ respectively. 
By applying the Stirling approximation for the microcanonical caliber
\begin{equation}\label{eq:stirling}
    S_{dt} = \ln \Omega_{\rm MB} \approx \sum_{ij} \left[n_{ji}\ln g_j - n_{ji} (\ln n_{ji} -1)\right],
\end{equation}
and the extremum condition $\partial f_{\rm MB} / \partial n_{ji} = 0$, we find
\begin{equation}\label{eq:mbnji}
    n_{ji} = g_j e^{-\beta \epsilon_j - \theta_i}.
\end{equation}
The constraint~Eq.~\eqref{eq:num} on $n_i(t)$ determines values of the Lagrange multipliers $\theta_i$:
\begin{equation}\label{eq:mbtheta}
    e^{-\theta_i} = \frac{n_i}{Z_{\rm MB}},
\end{equation}
in which $Z_{\rm MB} = \sum_j g_j \exp(-\beta\epsilon_j)$ is the canonical partition function. Combined, Eqs.~\eqref{eq:mbnji} and \eqref{eq:mbtheta} yield
\begin{equation}\label{eq:mbfinal}
    n_{ji} = \frac{g_j}{Z_{\rm MB}} e^{-\beta\epsilon_j} n_i = g_j \pi_{ji} n_i = p_{ji} n_i,
\end{equation}
where the elements $\pi_{ji} = \exp(-\beta\epsilon_j) / Z_{\rm MB}$, which in this simple model do not actually depend on the index $i$, express the probability for a particle to jump from an $i$\textsuperscript{th} energy level into \textit{one} of the $g_j$ equivalent sublevels of energy $\epsilon_j$. In addition, we can define the probability $p_{ji} = g_j \pi_{ji}$ for a particle to jump into \textit{any} of the degenerate sublevels $j$. Note here that $\pi_{ji}$ satisfy the simplest form of detailed balance, whereas $p_{ji}$ obey a more general relation:
\begin{equation}\label{eq:db}
\frac{\pi_{ji}}{\pi_{ij}} = e^{-\beta (\epsilon_j - \epsilon_i)},\qquad
\frac{p_{ji}}{p_{ij}} = \frac{g_j}{g_i} e^{-\beta \left(\epsilon_j - \epsilon_i\right)}.
\end{equation}

Using the above results and definitions, we can use the constraint Eq.~\eqref{eq:num} on $n_j(t+dt)$ as a Markov-chain equation for the most likely evolution---the macroscopic path---of the system:
\begin{equation}\label{eq:chain}
    n_j(t+dt) = \sum_i p_{ji} n_i(t).
\end{equation}
Due to the detailed-balance relations~\eqref{eq:db}, the system will reach a steady state $\bar{n}_j = n_j(\infty)$ as $t\to\infty$:
\begin{equation}\label{eq:micro}
    \bar{n}_j = \sum_i p_{ji} \bar{n}_i = g_j e^{-\beta\epsilon_j} \frac{N}{Z_{\rm MB}} = g_j e^{\beta(\mu_{\rm MB} - \epsilon_j)},
\end{equation}
which is the distribution of the microcanonical equilibrium ensemble with inverse temperature $\beta = (\kB T)^{-1}$ and chemical potential $\mu_{\rm MB}= -\kB T \ln (Z_{\rm MB} / N)$, cf. Eq.~\eqref{eq:ni_eq} in \apx~\ref{app:boltzmann}.

Instead of maximizing the Boltzmann entropy~(\apx~\ref{app:boltzmann}), we derived the distribution function~Eq.~\eqref{eq:micro} of the microcanonical ensemble alongside the Markovian macroscopic dynamics~Eq.~\eqref{eq:chain}, which encompasses the transient relaxation of any initial state $n_{i}(t)$ to equilibrium $\bar{n}_i(\infty)$. This theory can also be extended to continuous-time random walks in a straightforward manner, by casting Eq.~\eqref{eq:chain} as a master equation (\apx~\ref{app:master}) in the limit of $dt\to 0$ with transition rates
\begin{equation}\label{eq:rate}
    k_{ji} = \lim_{dt\to0} \frac{p_{ji}}{dt}.
\end{equation}

In the rest of the paper, we consider increasingly more complex microscopic constraints, which incorporate various important aspects of real physical systems neglected in the paradigmatic example of this section. Further developments follow relying as above on the maximization of the discrete microcanonical \textit{caliber}~$S_{dt}$, here taking the form of $S_{dt} = \ln\Omega_{\rm MB}(\bm{s}_{dt})$.

\subsection{\label{sec:fermi}Fermi gas}
Although quantum physics is beyond the scope of this paper, we consider the exclusion principle---leading to Fermi-Dirac statistics~\cite[Chapter 13]{Carter2001}---as a paradigmatic example of steric interparticle interactions in classical thermodynamic systems. Namely, we forbid two or more particles to occupy the same sublevel of a given energy $\epsilon_i$. This condition restricts the maximum occupation number of an $i$\textsuperscript{th} level by its degeneracy $n_i \le g_i$. 
Indeed, besides fermions in quantum physics, a variant of the exclusion principle often appears in models of classical statistical mechanics as a simplified representation of hard-core repulsion, e.g. in Flory-Huggins solution theory~\cite{Flory1942,Huggins1942,Weber2019}, \cite[Section 1.4]{safran2018statistical} or lattice Boltzmann methods~\cite{Katz1984,WolfGladrow2000}. When the degeneracy of an energy level is associated with the spatial localization of its sublevels, this type of interaction results in volume exclusion~\cite{Janowsky1992,Derrida1998,Golinelli2006,Chou2011,Hilrio2020,Elliott2025}. Such a situation may describe a collection of particles and boxes too small to host more than one particle. The Fermi-Dirac distribution arises through maximization of the Boltzmann entropy at constant number of particles and constant energy, by using the Stirling approximation.\footnote{The Stirling approximation requires both the number of particles in a given energy level $n_i$, as well as the number of unoccupied sublevels $g_i - n_i$ to be sufficiently large. Nonetheless, despite its asymptotic nature, the error in Stirling’s approximation is generally negligible on macroscopically relevant scales, even for small positive integers.}

As in Sec.~\ref{sec:ideal} the system's trajectory $\bm{s}_{dt}$ over a time step $dt$ is completely specified by its elements $n_{ji}$. However, while the particles are allowed to jump into any level (Fig.~\ref{fig:ideal}), in addition to constraints~\eqref{eq:num} and \eqref{eq:eng}, the exclusion principle demands
\begin{equation}\label{eq:exclusion}
    n_i(t) = \sum_j n_{ji} \le g_i,\quad
    n_j(t + dt) = \sum_i n_{ji} \le g_j.
\end{equation}
With the above listed assumptions, counting realizations of a path subject to the constraints~\eqref{eq:exclusion} leads to a formula of the system's caliber that resembles the Fermi-Dirac statistics (\apx~\ref{app:fd}):
\begin{equation}\label{eq:fd}
    \Omega_{\rm FD}(\bm{s}_{dt}) = \prod_j \frac{g_j!}{(g_j - \sum_i n_{ji})! \prod_i n_{ji}}.
\end{equation}
By maximizing $\ln \Omega_{\rm FD}$ with Lagrange multipliers $\beta$ and $\theta_i$ to the constraint Eqs.~\eqref{eq:num} and \eqref{eq:eng} we further obtain (\apx~\ref{app:fd})
\begin{equation}\label{eq:fdmax}
    n_{ji} = \frac{g_j - n_j'}{Z_{\rm FD}} e^{-\beta\epsilon_j} n_i = p_{ji} n_i,
\end{equation}
where we used for brevity $n_j' = n_j(t + dt)$ and, as in Sec.~\ref{sec:ideal}, the transition probabilities $p_{ji}$, which depend on the Fermi-Dirac partition function $Z_{\rm FD} = \sum_j \left(g_j - n_j'\right) \exp(-\beta\epsilon_j)$.
From the constraint Eq.~\eqref{eq:num} on $n_j' = \sum_i n_{ji}$ we again obtain a Markov chain
$$
    n_j(t + dt) = \sum_i p_{ji} n_i(t),
$$
whose steady state must satisfy the equation
\begin{equation}\label{eq:fdss}
    \bar{n}_j = \sum_i p_{ji} \bar{n}_i = \frac{g_j - \bar{n}_j}{Z_{\rm FD}} e^{-\beta\epsilon_j} N.
\end{equation}
The solution of Eq.~\eqref{eq:fdss} is the equilibrium Fermi-Dirac distribution with the chemical potential $\mu_{\rm FD} = -\kB T \ln(Z_{\rm FD}/N)$:
\begin{equation}\label{eq:fdeq}
    \bar{n}_j = \frac{g_j}{1 + e^{\beta(\epsilon_j - \mu_{\rm FD})}}.
\end{equation}

Note that the microcanonical caliber theory imposes a general condition on the transition probabilities $p_{ji}$:
\begin{equation}\label{eq:fddb1}
    \frac{p_{ji} \bigl(g_i - n_i(t + dt)\bigr)}{p_{ij} \bigl(g_j - n_j(t+dt)\bigr)} = e^{-\beta(\epsilon_j - \epsilon_i)},
\end{equation}
which in equilibrium reduces to that of the detailed balance for fermions~\cite{Bowlden1957,Nguyen2015}
\begin{equation}\label{eq:fddb2}
    \frac{p_{ji} (g_i - \bar{n}_i)}{p_{ij} (g_j - \bar{n}_j)} = e^{-\beta(\epsilon_j - \epsilon_i)}.
\end{equation}

The Fermi gas example shows how the microcanonical principle of maximum caliber reproduces the classical results of statistical mechanics for such systems, e.g. Eqs.~\eqref{eq:fdeq} and \eqref{eq:fddb2}, and provides their nonequilibrium generalizations, e.g. Eq.~\eqref{eq:fddb1} and those derived in the following. Interactions more complex than the exclusion principle can also be taken into account (\apx~\ref{app:inter}, Ref.~\cite{Elliott2025}).

\section{\label{sec:main}Spatially extended systems}
\citet[\S I,2]{GrootMazur} argue in their, now classical, book on nonequilibrium thermodynamics, which they regard in connection to such macroscopic disciplines as fluid dynamics and electromagnetic theory, that ``\textit{the thermodynamics of irreversible processes should be set up from the start as a continuum theory, treating the state parameters of the theory as field variables...}''

To pursue the above idea, we first reframe the Boltzmann model of the ideal gas (Sec.~\ref{sec:ideal}) using discretized field variables on a lattice. Then we analyze its thermodynamics in and out of equilibrium by applying the microcanonical principle of maximum caliber.

\begin{figure}[!t]\centering
\includegraphics[width=\columnwidth]{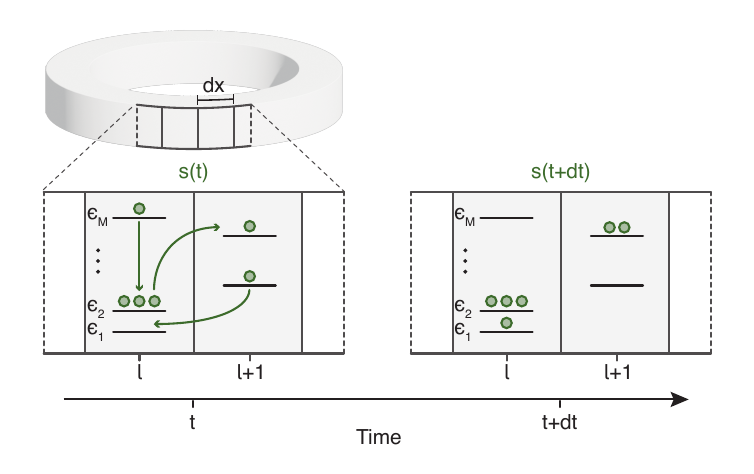}
\caption{\label{fig:lattice}Boltzmann gas on a lattice of length scale $dx$ with periodic boundary conditions. In general the structure of energy levels 
%$\epsilon_{i=1,2,...,M(\ell)}(\ell=1,2,...,L)$ 
$\epsilon_{i}(\ell)$ with $i = 1 , \dots, M(\ell)$ and $\ell=1,2,...,L$,
is different at each site of the lattice. In a time step $dt$ particles are allowed to jump only between the levels of the same site $\ell$ or of its nearest neighbors $\ell\pm1$.
}
\end{figure}

\subsection{\label{sec:eq}Boltzmann gas on a periodic lattice}
More realistic representations of macroscopic systems can be achieved by associating energy levels with sites on a lattice, e.g.\ on an equidistant grid of size $dx$ which in the continuum limit $dx\to0$ yields a field description extended over the coordinate $x$ (\apx~\ref{app:diffusion}). For example, such lattice sites may correspond to coarse-grained positions of molecules in space (Fig.~\ref{fig:lattice}). Here we consider the Boltzmann gas with energy sets 
$\epsilon_{i}(\ell)$ with $i = 1 , \dots, M(\ell)$ and $\ell=1,2,...,L$,
and degeneracies $g_i(\ell)$, where the index $\ell$ refers to an $\ell$\textsuperscript{th} site. Thereby the system represents a (quasi-) one-dimensional system with particles' energy states localized in space. In addition, we impose periodic boundary conditions. 

We introduce a new constraint on the microscopic dynamics, which corresponds to the continuity conditions of continuum fields, and which restricts possible transitions from a source level $\epsilon_i(\ell)$ to the target levels $\epsilon_j\bm{(}m \in \mathcal{N}(\ell)\bm{)}$ within the \textit{neighborhood} $\mathcal{N}(\ell) $ of the $\ell$\textsuperscript{th} site. We take the simplest such neighborhood $\mathcal{N}(\ell) = \{\ell - 1, \ell, \ell + 1\}$ with periodic boundary conditions $\mathcal{N}(1) = \{L, 1, 2\}$ and $\mathcal{N}(L) = \{L - 1, L, 1\}$.

Using two indices, instead of one, to enumerate the energy levels does not change much in the framework outlined in Sec.~\ref{sec:framework}. In fact, double indexing can be in practice ``flattened out,'' given a finite set $i \in \mathcal{M}(\ell)=\{1,2,...,M_\ell\}$ of energy levels $\epsilon_i(\ell)$ per site $\ell$. Hence the count of realizations of a path specified by $n(m j|\ell i)$---the number of particles transferred between the source and target levels $\epsilon_i(\ell)$ and $\epsilon_j(m)$, respectively---is still given by the Maxwell-Boltzmann statistics
\begin{equation}\label{eq:omega0}
    \Omega_{0}\left(\bm{s}_{dt}\right) \approx \prod_{\substack{\ell=1\\i\in\mathcal{M}(\ell)}}^L \prod_{\substack{m\in\mathcal{N}_\ell\\j\in\mathcal{M}(m)}}\frac{g_j(m)^{n(mj|\ell i)}}{n(mj|\ell i)!},
\end{equation}
cf. Eq.~\eqref{eq:mb}. We also impose the constraints
\begin{align}\label{eq:num0}
    A_{\ell i} = n_i(\ell, t) - \sum_{m\in\mathcal{N}(\ell),j\in\mathcal{M}(m)} n(m j| \ell i) = 0,&
    \\\label{eq:eng0}
    B = E - \sum_{\substack{\ell=1,m=\ell-1 \\ i\in\mathcal{M}(\ell),j\in\mathcal{M}(m)}}^{L,\ell+1} \epsilon_j(m) n(m j| \ell i) = 0,&
\end{align}
subject to the periodic boundary conditions $n_i(L + 1) = n_i(1)$, 
thus applying the condition of conservation of number of particles and energy expressed by Eqs.~\eqref{eq:num} and \eqref{eq:eng} to the case of transitions from a given source level to a neighborhood $\mathcal{N}(\ell)$.
Next, to maximize the system's caliber, we form the objective function \begin{equation}\label{eq:f0}
    f_0 = \ln \Omega_0(\bm{s}_{dt})+ \beta B(\bm{s}_{dt}) + \sum_{\ell i} \theta_{\ell i} A_{\ell i}(\bm{s}_{dt}),
\end{equation}
with the Lagrange multipliers $\theta_{\ell i}$ and $\beta$ to the constraints~\eqref{eq:num0} and \eqref{eq:eng0} respectively. We do not repeat here the steps of Sec.~\ref{sec:ideal} required to extremize $f_0$ and to eliminate $\theta_{\ell i}$, and write immediately the final result for the most likely path of the system
\begin{equation}\label{eq:path0}
    n(m j | \ell i) = \frac{g_j(m)}{\zeta(\ell)} e^{-\beta\epsilon_j(m)} n_i(\ell) = p(mj|\ell i) n_i(\ell),
\end{equation}
in which we introduced a \textit{neighborhood} partition function
$$
    \zeta(\ell) = \sum_{m\in\mathcal{N}(\ell), j\in\mathcal{M}(m)} g_j(m) e^{-\beta \epsilon_j(m)}.
$$
and the transition probabilities $p(mj|\ell i)$, with $m\in\mathcal{N}_\ell$. The latter can be augmented by elements $p\bm{(}m \not\in \mathcal{N}(\ell), j|\ell i\bm{)} = 0$ to cast a Markov chain equation, cf. Eq.~\eqref{eq:chain},
\begin{equation}\label{eq:chain0}
    n_j(m, t + dt) = \sum_{\ell i} p(mj|\ell i) n_i(\ell),
\end{equation}
in which we observe another generalization of the detailed-balance relation
\begin{equation}\label{eq:db0}
    \frac{p(mj|\ell i)}{p(\ell i|mj)} = \frac{g_j(m) \zeta(\ell)}{g_i(\ell) \zeta(m)} e^{-\beta\bigl(\epsilon_j(\ell) - \epsilon_i(m)\bigr)},
\end{equation}
when $m\in\mathcal{N}(\ell)$. In a uniform system with $\zeta(m) = \zeta(\ell)$ and $g_j(m) = g_i(\ell)$ this relation simplifies to the more common form.

The equilibrium state corresponds to the steady-state solution of Eq.~\eqref{eq:chain0}:
\begin{equation}\label{eq:eq}
    \bar{n}_j(m) = \sum_{\ell i} p(mj|\ell i) \bar{n}_i(\ell)
    = \frac{g_j(m)}{e^{\beta\epsilon_j(m)}} \sum_{\ell\in\mathcal{N}(m)}\frac{\bar{\nu}(\ell)}{\zeta(\ell)},
        %= g_j(m) e^{\beta\bigl(\bar{\mu}(m) - \epsilon_j(m)\bigr)}
\end{equation}
where $\nu_{\ell} = \sum_i n_i(\ell)$ counts the particles at the site $\ell$. This linear equation can be solved explicitly, once the structures of energy levels---how $g_i(\ell)$ and $\epsilon_i(\ell)$ depend on the indices $i$ and $\ell$---are defined. Inhomogeneous structures are possible, e.g.\ in the presence of external forces like gravity. In the absence of such forces the equilibrium solution manifests the macroscopic symmetry, which demands that all sites, as they are equivalent, are characterized by the same numbers
\begin{equation}\label{eq:uniform}
    \bar{\nu}(\ell) \equiv \frac{N}{L},\qquad
    \zeta(\ell) \equiv 3 z = 3 \sum_j g_j e^{-\beta\epsilon_j},
\end{equation}
where we drop the dependence of $g_j$ and $\epsilon_j$ on the index $m$. The factor of $3$ appears from expansion of the neighborhood partition function $\zeta$ comprising contributions of the local partition function $z$ from three identical lattice sites. Equation~\eqref{eq:uniform} substituted into \eqref{eq:eq} reproduces the macroscopic statistics Eq.~\eqref{eq:micro} for the Boltzmann gas:
\begin{equation}\label{eq:micro0}
    \bar{n}_j(m) = g_j e^{\beta\bigl(\mu_{\rm MB} - \epsilon_j\bigr)},
\end{equation}
in which the chemical potential at each site is given by
\begin{equation}\label{eq:mumb}
    \mu_{\rm MB} = \kB T \ln \frac{3 N/L}{3 z} = -\kB T \ln\frac{Z_{\rm MB}}{N},
\end{equation}
with $Z_{\rm MB} = L z$ being the macroscopic partition function.

The microcanonical principle of maximum caliber, as shown above, encompasses the Markovian dynamics~Eq.~\eqref{eq:chain0}, which describes the decay of the transients towards the 
%terminal 
equilibrium states. Thereby this approach consistently extends the ensemble formalism of classical statistical mechanics. In addition to the continuous-time limit, mentioned in Sec.~\ref{sec:ideal} and leading to the master equation (\apx~\ref{app:master}), the lattice representation connects the microcanonical principle of maximum caliber to the field theories through the continuum limit of $dx\to0$ (\apx~\ref{app:diffusion}).
% Moreover, comparison of Eqs.~\eqref{eq:eq} and \eqref{eq:mumb} suggests a definition of a chemical potential at a site $m$ for any state $\bm{s}$:
% \begin{equation}\label{eq:mu0}
%     \mu(m) = \kB T \ln \frac{3\nu(\ell)}{\zeta(\ell)}.
% \end{equation}
%which contains local contributions from $\nu(m)/\zeta(m)$, as well as nonlocal terms $\mu(m\pm1)/\zeta(m\pm1)$ introducing spatial correlations. These correlations disappear in equilibrium. %  As demonstrated in the following sections, this definition of chemical potential applies also to nonequilibrium steady states.

%%
\subsection{\label{sec:grad}Gradient-driven flow}
Steady states with constant macroscopic flows, which are sustained by external gradients of thermodynamic potentials, represent the classical subject of nonequilibrium thermodynamics, cf. Refs.~\cite[Chapter V]{GrootMazur} and \cite[Sec. 1.5]{Attard2012}. Here we show how the microcanonical-caliber framework can be applied to analyze statistical mechanics of such systems.

Extending the ideal-gas model on a lattice from Sec.~\ref{sec:eq}, we drive the system out of equilibrium by imposing a macroscopic constraint enforcing a constant flux of particles $J$ between the end sites $\ell = 1,L$:
\begin{equation}\label{eq:grad}
    P_{1L}(\bm{s}_{dt}) = J - \sum_{ij} \left[n(1j|Li) - n(Lj|1i)\right] = 0, 
\end{equation}
which implies $J > 0$ when the particles are driven counter-clockwise in Fig.~\ref{fig:lattice}. Note that this constraint is \textbf{not} equivalent to constant flux in the whole system, as clarified in the next section.

The nonequilibrium mechanism we thus describe resembles closing an electric circuit by connecting its ends to positive and negative terminals of a ``battery.'' Then, by adding the additional constraint $P_{1L}$ with a new Lagrange multplier $\eta_{1L}$ to Eq.~\eqref{eq:f0}, we extremize the objective function
\begin{equation}\label{eq:fgrad}
    f_{1L} = f_0 + \eta_{1L} P_{1L}.
\end{equation}
For $m,\ell \not\in\{1,2, L-1, L\}$ we obtain again Eq.~\eqref{eq:path0}, as these sites are not affected by the new constraint. The explicit expressions for the affected sites, which involve the Lagrange multiplier $\eta_{1L}$, are derived in \apx~\ref{app:grad} [Eq.~\eqref{eq:n1jLi} and \eqref{eq:nLj1i}].

The dynamics of the system can be cast as Eq.~\eqref{eq:chain0} by using~\eqref{eq:path0}, \eqref{eq:n1jLi}--\eqref{eq:nLj1i}, like in Sec.~\ref{sec:eq}. Due to the constant-flux condition, its steady-state solution $\bar{n}_i$ must obey Kirchhoff's law:
\begin{equation}\label{eq:kirchhoff}
    J_{\ell+1,\ell} = \sum_{ij} \left[n(\ell + 1,j|\ell i) - n(\ell,j|\ell+1,i)\right] = J.
\end{equation}

Now we focus on a subsystem consisting of sites $\ell\in S_0 = \{a, a+1, ..., b\}$ with $a\gg1$ and $b\ll L$. The complementary subsystem $\ell\in \bar{S}_0 = \{b + 1, b+2,...,a-1\}$, internally connected through the periodic boundary conditions, plays the role of a \textit{nonequilibrium} reservoir.

In the subsystem of interest $S_0$, Eq.~\eqref{eq:kirchhoff} can be evaluated with help of \eqref{eq:path0}:
$$J_{\ell+1,\ell}(S_0) = \frac{z(\ell+1)}{\zeta(\ell)} \bar{\nu}(\ell) - \frac{z(\ell)}{\zeta(\ell +1)} \bar{\nu}(\ell+1),$$
which for the Boltzmann gas with homogeneous structure of energy levels reduces to
$$
    J_{\ell+1,\ell}(S_0) = -\frac{z}{\zeta} \Delta\bar{\nu}(\ell) = -\frac{1}{3}\Delta\bar{\nu}(\ell) = J,
$$
with the forward difference $\Delta\bar{\nu}(\ell) = \bar{\nu}(\ell+1) - \bar{\nu}(\ell)$. Hence, the steady-state characterized by the constraint~\eqref{eq:grad} induces a constant gradient of matter in the opposite direction of the flux it sustains within the bulk of the system.

If we assume that the complementary system $\bar{S}_0$ acts as a perfect reservoir, it can be eliminated from the model by imposing on the subsystem of interest $S_0$ constant boundary conditions $\nu(\ell\in\{a - 1, b+1\}) = \bar{\nu}(\ell)$ at the sites $a - 1$ and $b + 1$ . In principle, given the fixed number of bins and the energy levels' structure, the Lagrange multiplier $\eta(J)$ and the gradient it induces can be evaluated at least numerically.

At the level of the whole system, the nonequilbrium steady state is sustained by a pair of asymmetric \textit{active exponents} $\phi_{1L} = -\phi_{L1} = -\eta_{1L}$ introduced in Ref.~\cite[Supplemental Material Sec. 2]{Belousov2024} and discussed in \apxs~\ref{app:grad} and \ref{app:active}. They break the detailed-balance relation for the transition rates and, instead of Eq.~\eqref{eq:db0}, yield
\begin{equation}\label{eq:dbphi}
    \frac{p(Lj|1i)}{p(1j|Li)} = \frac{g_j(L) \zeta_{\phi}(1)}{g_i(1) \zeta_\phi(L)} e^{-\beta\bigl(\epsilon_j(\ell) - \epsilon_i(m)\bigr) + \Delta\phi},
\end{equation}
where $\Delta{\phi} = \phi_{1L} - \phi_{L1} = - 2 \eta_{L1}$ and $\zeta_{\phi}(\ell \in \{1, L\})$ is the \textit{extended} neighborhood partition function (\apx~\ref{app:grad}). The active exponents, which here depend through $\eta_{1L}$ on both the flux $J$ and the system's state, effectively behave as feedback agents---or ``thermodynamic demons''---and exploit the information they have about the system to drive it perpetually out of equilibrium. In the next section we discuss a different mechanism sustaining a nonequilibrium steady state with a constant flux.

\subsection{\label{sec:act}Active directed and diffusive motion}
Active self-propelling particles constitute another class of nonequilibrium systems widely studied with the methods of statistical mechanics~\cite{Bechinger2016}. Here, we discuss two cases in which active particles move in a diffusive or a directed manner on a periodic lattice.

Each site of the whole lattice filled with self-propelling particles can be regarded as a subsystem. In the ensemble of such subsystems, the propensity of the particles to undergo directed active motion~\cite{Svetlov2023,Bishop2023} can be described by a fraction $\psi_+ > 1 / 3$ of their total count $N$, which in a time step $dt$ always jumps in the same direction from one site $\ell$ to the other $\ell+1$.
This mechanism may describe a limited amount of ``fuel'' $N(\phi_+ - 1/3)$ available to power the directed motion~\cite{Kavi2025} and thus bias the expected otherwise fraction of $1 / 3$ molecules moving in the chosen direction.

When $N, L\to\infty$, we expect to observe a total \textit{cumulative} current in the whole system $J_{\rm DM} = \psi_+ N$. Therefore, instead of the constraint Eq.~\eqref{eq:grad} we use
\begin{equation}\label{eq:hat}
    P_{\rm DM} = J_{\rm DM} - \sum_{\ell ji} n(\ell + 1, j|\ell i) = 0.
\end{equation}
Of course the direction of motion can be reversed if necessary.

By a similar token, active or enhanced diffusion~\cite{Abbaspour2021,Meng2023} can be implemented by requiring a certain fraction of particles to jump to either side of the source site $\ell$, rendering a cumulative displacement $J_{\rm AD}$:
\begin{equation}\label{eq:tilde}
    P_{\rm AD} = J_{\rm AD} - \sum_{\ell ji} \bigl(n(\ell - 1, j|\ell i) + n(\ell + 1, j|\ell i)\bigr).
\end{equation}
By using the Lagrange multipliers $\eta_{\rm DM}$ and $\eta_{\rm AD}$ to the constraints $P_{\rm DM}$ and $P_{\rm AD}$, respectively, we form the objective functions
\begin{align}\label{eq:fhat}
    f_{\rm DM} =& f_0(\bm{s}_{dt}) + \eta_{\rm DM} P_{\rm DM}(\bm{s}_{dt}),\\\label{eq:ftilde}
    f_{\rm AD} =& f_0(\bm{s}_{dt}) + \eta_{\rm AD} P_{\rm AD}(\bm{s}_{dt}).
\end{align}
By extremizing Eqs.~\eqref{eq:fhat} and \eqref{eq:tilde} we obtain the most likely paths
\begin{align}\label{eq:nhat}
    n_{\rm DM}(mj|li) =& g_j e^{-\beta\epsilon_j} \frac{n_i}{\zeta_\phi(\ell)} \exp(\hat\phi_{m\ell}),
    \\\label{eq:ntilde}
    n_{\rm AD}(mj|li) =& g_j e^{-\beta\epsilon_j} \frac{n_i}{\zeta_\phi(\ell)} \exp(\tilde\phi_{m\ell}),
\end{align}
which are expressed here by using active exponents
\begin{align}\label{eq:phi}
    \hat\phi_{m\ell} =& \begin{cases}
        -\eta_{\rm DM}\text{ if $m = \ell + 1$,}\\
        0\text{ otherwise,}
    \end{cases}\\
    \tilde\phi_{m\ell} =& \begin{cases}
        -\eta_{\rm AD}\text{ if $m \ne \ell$,}\\
        0\text{ otherwise,}
    \end{cases}
\end{align}
cf.\ \apxs~\ref{app:grad} and \ref{app:active}. We omit the derivation of the steady states in the active systems considered here, as it repeats the steps covered earlier in a straightforward manner.

In a system with a homogeneous structure of energy levels, all sites must be macroscopically equivalent. Therefore, due to symmetry considerations, the nonequilibrium conditions do not generate a gradient of particles in the steady state, even in the presence of a flux $J_{\rm DM}$. 
Unlike the case of gradient-driven flow discussed in the previous section, here the active exponents~\eqref{eq:phi} also contain a symmetric part---called \textit{frenetic} in Ref.~\cite{Maes2020}--- 
i.e.\ $\hat\phi_{\ell,\ell+1} + \hat\phi_{\ell+1,\ell} \not\equiv 0$ and $\tilde\phi_{\ell,\ell+1} + \tilde\phi_{\ell+1,\ell} \not\equiv 0$. This symmetric part gives rise to \textit{active} fluctuations~\cite[Supplemental Material Sec. 2]{Belousov2024}, as well as to the notions of effective temperature and force-dependent transport coefficients\footnote{here synonymous to nonlinear constitutive relations} (\apx~\ref{app:active}).

\subsection{\label{sec:ensembles} Norton and Th\'evenin ensembles}
Sec.~\ref{sec:grad} describes a \textit{microcanonical Norton ensemble} with a fixed current at the boundaries---a method sometimes invoked in nonequilibrium atomistic simulations~\cite{Evans1986,MllerPlathe1999}, \cite[Sec. 6.6]{EvansMorris2007}. There exists a complementary approach, Th\'evenin ensemble~\cite[Sec. 6.6]{EvansMorris2007}, which leverages the concept of ensemble equivalence to fix, instead of the current $J$, its conjugate potential~\cite{Miller1979,Kato2001}. Such ideas have already been put into practice, e.g.\ to model heat flow~\cite{Conti2013}.

In Sec.~\ref{sec:grad} the microcanonical principle of maximum caliber identifies the Lagrange multiplier~$\eta$---called \textit{current affinity} in the following---as the conjugate variable of the boundary flux $J$ that does not correspond precisely to the phenomenological quantities found in the literature~\cite{Miller1979,Kato2001,Conti2013}. By using our approach, we can analyze how and when the duality between a current and its affinity allow the equivalence of Norton and Th\'evenin ensembles (Sec.~\ref{sec:num}).

As a counter example we also consider below the directed motion from Sec.~\ref{sec:act}, since both the directed motion and active diffusion can be classified as Norton ensembles with the cumulative flux $J_{\rm DM}$ and displacement $J_{\rm AD}$ as the generalized currents. However, as we discuss further, a fixed affinity $\eta_{\rm DM}$ does not generate the same statistics of the macroscopic flow as the constrained current $J_{\rm DM}$ in this context.

Indeed, the current affinity $\eta < 0$, which is conjugate to the boundary flux $J$, increases the fraction of particles that jump from the site $L$ to $1$, cf. Eq.~\eqref{eq:nLj1i} with $\phi_{L1} = - \eta$.
Likewise, it decreases the fraction of transitions in the reverse direction. Given that the steady state is on average characterized by a constant number of particles $\bar{\nu}_\ell$ with a global gradient $\bar\nu_{\ell + 1} - \nu_\ell \ne 0$, the biases of transition probabilities $p_{1L}$ and $p_{L1}$ sustain the constant flow $J = p_{1L} \bar\nu_L - p_{L1}\bar\nu_{L1}$. Therefore, in the steady state the current affinity, which in the Norton ensemble is a thermodynamic observable, should also attain its stationary value $\avg{\eta}$ in correspondence to $\bar{\nu}_1$ and $\bar\nu_L$.

If instead of the current $J_{1L}$ we fix the current affinity at the above value $\avg{\eta} < 0$, we can generate a microcanonical Th\'evenin ensemble, whose steady state must also be characterized by a gradient, as well as by a fluctuating flux of the mean value $\avg{J_{1L}}$ that compensates for the bias in $p_{1L}$ and $p_{L1}$ by $p_{1L}(\eta) \bar\nu_{1L} - p_{L1}(\eta) \bar{\nu}_{L1} - \avg{J_{1L}} = 0$. Here all the parameter values must be equal to those of the steady-state Norton ensemble. However, in the bulk of these two systems, dynamics is driven by the same transition probabilities. In absence of long-range correlations, an example of which we discuss shortly, the effect of fluctuations in the current affinity $\eta$ about its mean $\avg{\eta}$ or in the current $J_{1L}$ should be negligible away from the boundaries. Therefore in the steady state both ensembles should be characterized by the same average values of thermodynamic variables, when measured in the bulk of the system. Hence the Norton and Th\'evenin pictures are equivalent in this scenario.

However the directed motion formulated in Sec.~\ref{sec:act} creates long-range correlations across the whole system and thus destroys the equivalence of Norton and Th\'evenin ensembles. In particular, assuming the homogeneous structure of energy levels, the variance of the total flux $J_+ = \sum_\ell n_{(\ell+1)\ell} = J_{\rm DM}$, being the fixed quantity, must vanish: 
\begin{multline}\label{eq:correlation}
    \Var(J_+) = L \Var(n_{(\ell+1)\ell}) \\+ \sum_{\ell m} \operatorname{Cov}(n_{(\ell+1)\ell}, n_{(m+1)m}) = 0,
\end{multline}
which implies that covariances $\operatorname{Cov}(n_{(\ell+1)\ell}, n_{(m+1)m})$ between the numbers of transferred particles $n_{(\ell+1)\ell}$ and $n_{(m+1)m}$ all add up to a negative value to cancel the term $L \Var(n_{(\ell+1)\ell}) > 0$ in Eq.~\eqref{eq:correlation}.

In other words, the currents between \textit{all} sites of the lattice in the directed motion are negatively correlated: more particles moving to the right from one site imply fewer particles moving to the right from another site, and vice versa. This anticorrelation would be totally lost if we fix $\eta_{\rm DM}$ instead of the cumulative current $J_{\rm DM}$. In fact, this latter scenario would generate a steady state with the fraction of transitions to the right fixed \textit{independently} at each site. Depending on the system, such a constraint may be of interest, e.g.\ in the limit where the ``fuel'' powering the directed motion is in abundance~\cite{Kavi2025} or scales as $L\to\infty$ in the limit of a large system.

An important consequence of the anticorrelation Eq.~\eqref{eq:correlation} is suppression of current fluctuations demonstrated numerically in the following section. By losing this long-range correlation effect, a fixed affinity $\eta_{\rm DM} = -\hat{\phi}$ actually amplifies current fluctuations in the system by a factor of $3 [2 + \exp(-\hat\phi / 2)]^{-1} > 1$ (\apx~\ref{app:active}).

\begin{figure}[!t]\centering
\includegraphics[width=\columnwidth]{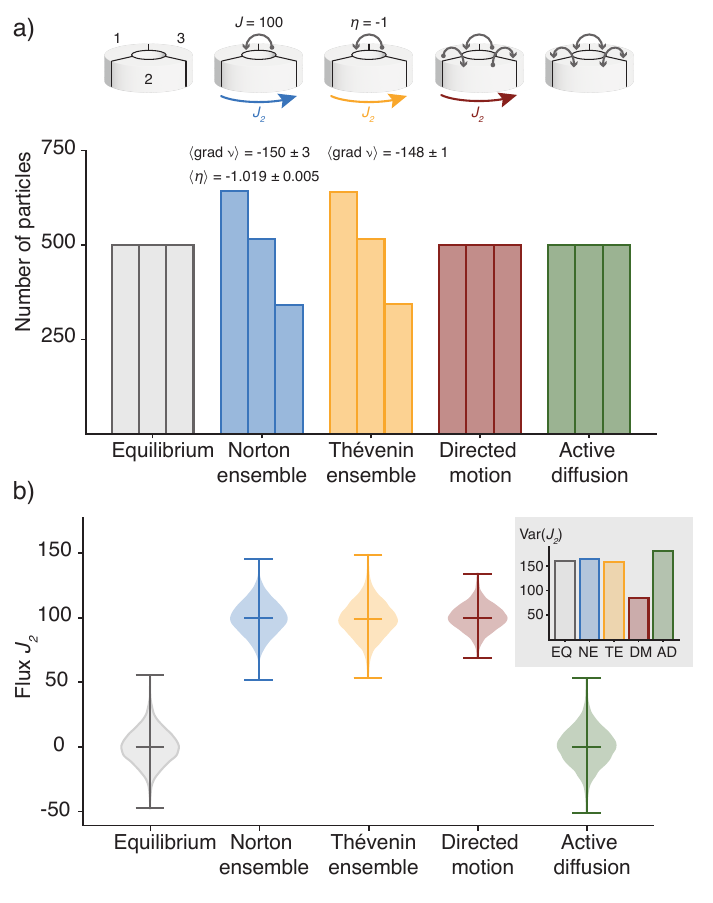}
\caption{\label{fig:sim}Diffusion of $N=1500$ particles on a periodic lattice of size $L=3$ under various conditions: \ref{it:eq}. Equilibrium conditions (EQ); \ref{it:grad}. Norton ensemble (NE) with a fixed boundary flux $J = 100$; \ref{it:ne}. Th\'evenin ensemble (TE) with a fixed current affinity $\eta = -1$; \ref{it:dir}. Directed motion (DM) with a fixed cumulative current $J_{\rm DM} = 700$; \ref{it:act} Active diffusion (AD) with a fixed cumulative displacement $J_{\rm AD} = 1100$. Panel (a): steady-state distributions of particles over lattice sites. Top icons summarize the simulation conditions. Panel (b): Distributions of the particles' current $J_2$ through the site $2$. The inset compares the variances $\Var(J_2)$ between different ensembles. Stochastic simulations of the Markov chain~Eq.~\eqref{eq:markov} are performed for $10100$ steps starting from a uniform distribution $\nu_i = N / L$ by applying a Gillespie-like algorithm (\apx~\ref{app:algo}). Measurements are taken over the last $10000$ simulation steps undersampled to $1000$ in order to ensure relaxation of the transient correlations. Standard errors correspond to five standard deviations of the mean.}
\end{figure}

%%%
\section{\label{sec:num}Numerical examples}
%%%
To illustrate the theoretical results of Sec.~\ref{sec:main}, we report below numerical simulations of a toy model on a small lattice of size $L = 3$~\cite{code}. Specifically, we consider the Boltzmann gas with a homogeneous structure of energy levels characterized by the same local partition function $z(\ell \in \{1,2,3\}) = z$. Furthermore, by summing out all the energy levels through the coarse-graining procedure described in \apx~\ref{app:diffusion}, we reduce the Markov chain~\eqref{eq:chain0} to
\begin{equation}\label{eq:markov}
    \bm{\nu}(t + dt) = \bm{p}\cdot\bm{\nu}(t),
\end{equation}
in which $\bm{\nu} = (\nu_1, \nu_2, \nu_3)$ is the vector of occupation numbers at each lattice site and $\bm{p}$ is the transition matrix. We analyze five distinct steady states to discuss several different features of nonequilibrium ensembles:
\begin{enumerate}
    \item \label{it:eq} As a baseline of comparison we consider equilibrium (Sec.~\ref{sec:eq}), in which all elements of the transition matrix $p_{\ell m}\equiv z / \zeta = 1/3$ are equal.
    \item \label{it:grad} Diffusion driven by a constant boundary flux $J$ (Sec.~\ref{sec:grad}) is the paradigmatic example of a system with a spatial gradient and flow of matter driven by a matrix
    \begin{equation}\label{eq:peta}
    \bm{p}(\eta) = \begin{pmatrix}
        (2 + e^{\eta})^{-1}
            & 1/3
            & (2 e^{\eta} + 1)^{-1}\\
        (2 + e^{\eta})^{-1}
            & 1/3
            & (2 + e^{-\eta})^{-1}\\
        (2 e^{-\eta} + 1)^{-1}
            & 1/3
            & (2 + e^{-\eta})^{-1}
    \end{pmatrix}
    \end{equation}
    parameterized by the current affinity $\eta$.
    \item \label{it:ne} Diffusion driven by a constant current affinity $\eta\equiv\const$, which fixes $\bm{p}$ through Eq.~\eqref{eq:peta}, illustrates the concept of equivalence between nonequilibrium ensembles.
    \item \label{it:dir} Active self propulsion with a fixed cumulative current $J_{\rm DM}$ sustained by a matrix $\bm{p}(\eta_{\rm DM})$ with elements
    \begin{equation}\label{eq:dir}
        p_{ij} = \begin{cases}
            (1 + 2 e^{\eta_{\rm DM}})^{-1}\text{ if $i = j + 1$},\\
            (2 + e^{-\eta_{\rm DM}})^{-1}\text{ otherwise},
        \end{cases}
    \end{equation}
    parametrized by the current affinity $\eta_{\rm DM}$. This case provides an example of matter flow in absence of a spatial gradient and with suppressed level of current fluctuations.
    \item \label{it:act} Active diffusion with a fixed cumulative displacement $J_{\rm AD}$ generated by a matrix $\bm{p}(\eta_{\rm AD})$ with elements
    \begin{equation}
        p_{ij} = \begin{cases}
            (2 + e^{\eta_{\rm DM}})^{-1}\text{ if $i \ne j$},\\
            (1 + 2 e^{-\eta_{\rm DM}})^{-1}\text{ otherwise},
        \end{cases}
    \end{equation}
    which entails neither spatial gradient nor macroscopic flow of matter, but amplifies microscopic fluctuations of the latter.
\end{enumerate}

To perform \textit{stochastic} simulations of steady-states~\ref{it:grad}, \ref{it:dir}, and \ref{it:act}, we designed an algorithm (\apx~\ref{app:algo}), which ensures the apposite constraints in each realization of the process $\bm{\nu}(t)$~\cite{code}. The Lagrange multipliers $\eta$, $\eta_{\rm DM}$, $\eta_{\rm AD}$ become observable random variates. Our algorithm is also applicable to lattices of sizes $L > 3$, but for conspicuity, we present the results of the minimal model with $L = 3$---the smallest size that, under periodic boundary conditions, allows a division into subsystems $S_0$ and $\bar{S}_0$ as described in Sec.~\ref{sec:grad}.

In our simulations we compute averages of the occupation numbers $\avg{\bm{\nu}}$, the gradient $
    \avg{\operatorname{grad} \nu} = \avg{\nu_3 - \nu_1} / 2
$, and the macroscopic flux $\avg{J_2} = \avg{n_{21} - n_{12}}$ through the lattice site $\ell = 2$ together with its variance $\Var(J_2)$ over a single long trajectory. Here $n_{m\ell}$ is the count of particles transferred from the $\ell$\textsuperscript{th} site to the $m$\textsuperscript{th} in a single step. In addition, in case~\ref{it:grad} we measure the current affinity $\avg{\eta}$, whose average value is then used to simulate case~\ref{it:ne}.

Because any transitions of particles between the three lattice sites are allowed, all simulations converge to steady states rather rapidly [Fig.~\ref{fig:sim}(a)]. For the equilibrium case~\ref{it:eq} ($\avg{J_2} \equiv 0$), we observe a flat uniform distribution $\nu = N / L$ ($\avg{\operatorname{grad}\nu}\equiv0$). However, this same profile corresponds to the nonequilibrium steady states of directed motion~\ref{it:dir} and active diffusion~\ref{it:act} as well.
In contrast, the constraint of a constant boundary flux generates a steady state with a gradient within the system, as we predict (Sec.~\ref{sec:grad}). By Kirchhoff's law the averaged flux in the central region of the system satisfies the equality $\avg{J_2} = J$.
We also verify the equivalence of Norton and Th\'evenin nonequilibrium ensembles \ref{it:grad} and \ref{it:ne}, respectively. The constant current affinity $\eta$, set in case~\ref{it:ne} at the average value observed in the Norton ensemble \ref{it:grad}, generates within the statistical uncertainties the current $\avg{J_2} = J$ and the gradient $\avg{\operatorname{grad}\nu}$ of the same magnitude and direction.

Our simulations clearly reveal that the absence or presence of the current in the system does not completely determine the statistical signatures of the steady-state ensemble. In cases \ref{it:eq}--\ref{it:ne} currents of vanishing or nonzero magnitude are characterized by the same amplitude of fluctuations [Fig.~\ref{fig:sim}(b)], but different distributions of particles. The directed motion of particles, case \ref{it:dir}, generates the same macroscopic flow $\avg{J_2}$ as in cases \ref{it:grad} and \ref{it:ne}, but the amplitude of its fluctuations is even lower than in equilibrium. Finally, a uniform profile of particles is observed in equilibrium \ref{it:eq}, steady states \ref{it:dir} and \ref{it:act}, but the mean value and variance of the flow $J_2$ differ across all these three cases.

The variance $\Var(J_2)$ differs between cases \ref{it:eq}--\ref{it:ne} and \ref{it:dir}--\ref{it:act}. In the context of case \ref{it:act} the active exponents contain a nonvanishing symmetric---frenetic---part $\phi_{m\ell} + \phi_{\ell m} \not\equiv 0$, which affects the amplitude of fluctuations in the system (\apx~\ref{app:active}). When this part is positive or negative, the level of stochastic noise in the system can be amplified or suppressed, respectively. In the case $\ref{it:dir}$, however, the cause of suppressed noise is the anitcorrelation constraint~Eq.~\eqref{eq:correlation} that reduces the fluctuations of the currents $n_{12}$, $n_{23}$ and $n_{31}$.

%%%
\section{Discussion}

Sections~\ref{sec:framework} and \ref{sec:main} outline a general theory of microcanonical caliber for a large class of systems. While further details of this theory and its specific applications require separate dedicated studies, we highlight areas in which our framework provides new theoretical insights.

\paragraph*{Beyond detailed-balance relations.}---To account for spontaneous fluctuations, which play an important role at meso- and microscopic scales, stochastic models of small thermodynamic systems are usually designed so as to observe the detailed-balance condition in the limit of vanishing nonequilibrium forces~\cite{Pradhan2010,Seifert2012,Maes2021,Maes2020,Freitas2021}. For example, in the context of master equations and discrete state spaces it is typically expressed as a ratio of the forward transition rates $k_{ji}$ from a state $\bm{s}_i$ to $\bm{s}_j$ and its reverse $k_{ji}$~\cite[Chapter 3]{PelitiPigolotti2021}:
\begin{equation}\label{eq:thermo}
    \frac{k_{ji}}{k_{ij}} = e^{-\beta \bigl(F(\bm{s}_j) - F(\bm{s}_i) + 
    %\beta
    \Delta{W}\bigr)},
\end{equation}
where $F(\bm{s})$ is a free energy of a state $\bm{s}$ and $\Delta{W}$ is the work done on the system by nonequilibrium forces. Indeed, the above equation is at the foundations of the stochastic-thermodynamics formalism~\cite{PelitiPigolotti2021,Maes2020}.

How general is Eq.~\eqref{eq:thermo} in nonequilibrium systems? The condition of detailed balance can be related to the first principles of classical microscopic dynamics~\cite{Evans2004I,Evans2004II,Evans2005,Maes2021}, although generalizations due to complex interactions exist, e.g. in Fermi-Dirac systems~\cite{Bowlden1957,Nguyen2015} (Sec.~\ref{sec:fermi}). This latter relation~\eqref{eq:fddb2}, as well as further generalizations~\eqref{eq:db}, \eqref{eq:db0}, \eqref{eq:fddb1}, \eqref{eq:dbphi}, and \eqref{eq:rw}, were routinely deduced here from the single principle of maximum caliber.

Indeed the microcanonical principle of maximum caliber provides a systematic means to derive and study the relation~\eqref{eq:thermo} with its generalization by \textit{direct calculations}. In special cases Eq.~\eqref{eq:thermo} may emerge under coarse graining like Eq.~\eqref{eq:rw}, in the mean-field approximation, or by taking the limit of Eq.~\eqref{eq:db0} with $g_i(\ell) = O\bigl(g_j(m)\bigr)$ and $\zeta(\ell) = O\bigl(\zeta(m)\bigr)$.

\paragraph*{Fluctuations in nonequilibrium systems.}---As shown in Secs.~\ref{sec:grad} and \ref{sec:act}, distinct nonequilibrium states can be generated by applying active exponents $\phi_{ij}$, which break the detailed balance. Our framework relates the statistics of fluctuating variables in such systems to constraints imposed on the microscopic dynamics, revealing how different driving mechanisms produce distinct statistical signatures, as discussed in Sec.~\ref{sec:num}. These mechanisms may entail generalized forms of the fluctuation-dissipation theorem involving effective temperature and force-dependent transport coefficients (\apx~\ref{app:active}). Therefore the microcanonical principle of maximum caliber also provides means to analyze the effect of nonequilibrium forces on the spontaneous fluctuations in thermodynamic systems.

\paragraph*{Space-dependent diffusion.}---External fields, such as chemical interactions at phase interfaces, sometimes cause two effects simultaneously: an effective thermodynamic potential $U(x)$ and a space-dependent diffusion coefficient $D(x)$ that appear in the Smoluchowski equation~\eqref{eq:SE} as independent parameters~\cite{Belousov2020,Belousov2022}. In the simplified model of Sec.~\ref{sec:eq}, such external fields could be described by a spatially inhomogeneous structure of energy levels summarized by neighborhood and local partition functions, $\zeta(\ell)$ and $z(\ell)$, which simultaneously generate a space-dependent diffusion coefficient $D(x)$ and an effective potential $U(x)$ derived in \apx~\ref{app:diffusion}. Thereby the microcanonical principle of maximum caliber reduces the two parameters of the Smoluchowski equation to a single origin and unifies phenomena of inhomogenous energetic and transport properties in thermodynamic systems.

\paragraph*{General transport processes.}---Unlike the original works of \citet{Filyukov1967I}, this paper develops the framework of microcanonical caliber for systems involving transport of matter, which is perhaps the simplest nonequilibrium phenomenon. Other processes frequently considered in the theoretical research on the maximum-caliber principle are heat and momentum transport, e.g.\ Refs.~\cite{Filyukov1967I,Filyukov1967II,Filyukov1968,Evans2004I,Evans2004II,Evans2005,Monthus2011}.

From the microcanonical perspective explored here, the theory of heat transport is more complicated, because to specify a system's path, in addition to the number of particles transferred between the energy levels, e.g. $n_{ji}$ in Sec.~\ref{sec:ideal}, one must also specify how the energy is being redistributed. Given that the energies of the source and target levels are in general not equal $\Delta\epsilon = \epsilon_j - \epsilon_i \ne 0$, additional constraints on the microscopic dynamics are necessary. For example, a transfer of particles $n_{ji}$ must be accompanied by another transfer $n_{m\ell}$ to compensate $\epsilon_m - \epsilon_\ell = -\Delta\epsilon$. The Lagrange multiplier $\beta$ maintains the global energy balance ensuring that all such changes everywhere in the systems add up to zero. Thereby the specific information about how much energy a particular transition $n_{m\ell}$ donates to or receives from another transition $n_{ji}$ is lost. In spatially extended systems, considerations of the macroscopic continuity equation for the energy, similar to Eq.~\eqref{eq:num0} for the number of particles, also come into focus, as the donor and acceptor transitions should reside within some neighborhood of each other.

The linear momentum is an example of another thermodynamic variable that can be introduced into the framework and accounted for separately to model related transport processes. At the level of microscopic dynamics additional constraints are required as well.

The above details, which are bypassed in the traditional approach to the maximum-caliber principle using the concept of thermodynamic reservoirs, should be explicitly developed at the level of microcanonical theory---a viable program that can now be pursued.

\paragraph*{First principles of microscopic dynamics.}---Ensemble theory is a cornerstone of statistical mechanics~\cite{Gibbs1902,tolman,gallavotti,Tuckerman2023}, rooted in the Hamiltonian deterministic dynamics of equilibrium systems, and more recently extended also to nonhamiltonian dynamics of dissipative systems \cite{j2007statistical,ECM2,evans2016,Caruso2020}. This theory provides the foundations of molecular-dynamics methods and guides the development of algorithms for nonequilibrium simulations~\cite{Tuckerman2023,EvansMorris2007,hoover1986nonequilibrium}. In fact, the theory of ensembles was developed by Boltzmann with the intent of describing the thermodynamic behavior of macroscopic systems, in the limit in which microscopic fluctuations are not observable. However, it turns out to be particularly useful in present-day science and technology dealing with small systems, in regimes where fluctuations cannot be neglected.

The mechanism of nonequilibrium driving introduced in Sec.~\ref{sec:grad} resembles the approaches developed in Refs.~\cite{Hafskjold1993,MllerPlathe1997,MllerPlathe1999} for atomistic modeling. Therefore, a connection between the microcanonical caliber theory and dynamical-systems formalism might be quite fruitful.

Our framework relies on the identification of key constraints for microscopic dynamics, which capture the essential properties of the physical systems being modeled. Due to the deterministic nature of classical physics, given coordinates $\bm{q}(t)$ and velocities $\bm{\dot{q}}(t)$ of the particles, there is in general a single dynamical constraint corresponding to Newton's second law
\begin{equation}\label{eq:newton}
    \begin{cases}
    \bm{q}(t + dt) = \bm{q}(t) + dt \bm{\dot{q}},\\
    \bm{\dot{q}}(t + dt) = \bm{\dot{q}}(t) + \frac{dt}{m} \bm{F}(\bm{q}),
    \end{cases}
\end{equation}
where we assume conservative forces $\bm{F}(\bm{q}) = \partial H / \partial\bm{q}$ within the microcanonical approach, and equal masses of particles $m$.

Once the initial state $\bm{s}(0) = \bigl(\bm{q}(0),\bm{\dot{q}}(0)\bigr)$ is specified, only one realization of the system's path $\bm{s}_t = \bigl(\bm{q}(t), \bm{\dot{q}}(t)\bigr)$ is possible, which can be formally written using the multidimensional delta function:
\begin{multline}\label{eq:lagrange}
    \Omega_\mathcal{T}\bigl(\bm{s}(0)\bigr) \propto \int_{\bm{\sigma} \times \mathcal{T}} \mathcal{D}\bm{s}_t\,
    \int_{\bm{\sigma}} \bm{ds}\,\bm{\delta}\bigl(\bm{s} - \bm{s}(0)\bigr)\,
    \\\times
    \bm{\delta}\left(
        \int_\mathcal{T} dt\, \frac{\delta\mathcal{L}}{\delta \bm{q}(t)}
    \right),
\end{multline}
where $\mathcal{L}$ is the system's \textit{Lagrangian} concisely summarizing Eq.~\eqref{eq:newton}. If the system is subject to periodic boundary conditions, the Lagrangian formalism can incorporate additional constraints and forces, such as those generating nonequilibrium driving in Sec.~\ref{sec:main}. For example, one can add a constant force on all particles together with a constraint of conserved energy to produce active flow as in Sec.~\ref{sec:act}.

In the context of dynamical systems, the uncertainty addressed by statistical methods may arise due to unknown initial-value conditions $\bm{s}(0)$, which can be characterized by a probability distribution $\psigma\bigl(\bm{s}(0)\bigr)$. This probability can replace the first delta function in Eq.~\eqref{eq:lagrange}---formally by performing a convolution $\psigma * \Omega_{\mathcal{T}}\bigl(s(0)\bigr)$. % Alternatively, it may be of interest to study the frequency of a given state $\bm{s}\in\bm{\sigma}$ in the long time evolution of a system.

\paragraph*{Conclusion.}---The microcanonical approach brings various physical insights, which can be applied well beyond the scope of this work, including information-theoretic tools developed for the maximum-caliber principle. These insights can also be combined with the concepts of canonical and other ensembles through the usual thermodynamic techniques~\cite{Tuckerman2023,Pearson1985}. In fact, Sec.~\ref{sec:num} explores one such example---the equivalence of the Norton and Th\'evenin ensembles. Finally the principle of maximum caliber may serve as a platform for unifying different theoretical approaches in statistical mechanics, such as random walks and dynamical systems.

\begin{acknowledgments}
R.B., J.E., and A.E. acknowledge funding from the EMBL. L.R. is members of the Gruppo Nazionale per la Fisica Matematica (GNFM) of Istituto Nazionale di Alta Matematica.
LR acknowledges funding from EMBL and is grateful for the work hospitality received at EMBL during a sabbatical visit. L.R. also acknowledges from Cascade funding calls of NODES Program, supported by the MUR---M4C2 1.5 of PNRR funded by the European Union---NextGenerationEU (Grant agreement no.\ ECS00000036).
The authors are grateful to Laeschkir W\"urthner, Patrick Jentsch, Pamela Guruciaga, Ian Estabrook, Tim Dullweber, and Johannes Jung for stimulating discussions.
\end{acknowledgments}

\appendix
\section*{Appendix}
\subsection{\label{app:variation}Maximum entropy and maximium caliber}
Equation~\eqref{eq:jaynes} defining the system caliber owes its inception to a peculiar mathematical property it shares with (Shannon-)Gibbs entropy Eq.~\eqref{eq:shannon}, when regarded as a functional of probability distribution functions~\cite{Haken1985,Haken1986I,Dewar2003,Dewar2005,Smith2011,Press2013}. If we impose constraints on some macroscopic variables $\avg{c_{i=0,1,2,...}(\bm{s})} = \bar{c}_i$ with known values $\bar{c}_i$, including the normalization condition $\int_{\bm{\sigma}} \bm{ds}\,\psigma(\bm{s}) = 1$ as $c_0(\bm{s}) \equiv 1$, and maximize $S_\sigma$ with respect to $\psigma(\bm{s})$, using the Lagrange multipliers $\lambda_{i=0,1,...}$, we obtain (see below)
\begin{equation}\label{eq:psigma}
    \psigma(\bm{s}) \propto \exp\left(-\sum_{i=0,1,2,...} \lambda_i c_i(\bm{s})\right).
\end{equation}
For example, if we choose $c_1(\bm{s}) = E(\bm{s})$ to be the system's energy, then Eq.~\eqref{eq:psigma} yields the canonical distribution $\psigma(\bm{s})=Z^{-1} \exp[-E(\bm{s})/(k_B T)]$ with the partition function $Z = e^{1 + \lambda_0}$, and temperature $T = (\kB\lambda_1)^{-1}$.

To maximize the Shannon-Gibbs entropy~\eqref{eq:shannon} with respect to to the distribution $\psigma(\bm{s})$ subject to the constraints on $\avg{c_{i=0,1,2}(\bm{s})} = \bar{c}_i$, we form an objective functional including Lagrange multipliers $\lambda_i$:
\begin{multline}
    F[\psigma(\bm{s})] = S_{\sigma} + \sum_i \lambda_i \bigl(\bar{c}_i - \avg{c_i(\bm{s})}\bigr) \\
    = - \int_{\bm{\sigma}} ds\, \psigma(\bm{s}) \left[
        \ln\psigma(\bm{s}) + \sum_i \lambda_i c_i(\bm{s})
    \right] + \sum_i \lambda_i \bar{c}_i.
\end{multline}
The variation of the objective function yields
$$
    \frac{\delta F}{\delta \psigma(\bm{s})} =  -\left(\ln\psigma(\bm{s}) + 1 + \sum_i \lambda_i c_i(\bm{s})\right).
$$
Imposing the condition of extremum $\delta F/\delta\psigma(\bm{s}) = 0$, we obtain
\begin{equation}
    \psigma(\bm{s}) = e^{- 1 - \lambda_i \sum_i c_i(\bm{s})}
        = Z_\sigma^{-1} e^{-\sum_{i>0} \lambda_i c_i(\bm{s})},
\end{equation}
in which we expanded the constraint $c_0(\bm{s}) \equiv 1$ of the normalized probability density $\bar{c}_0 = 1$ and introduced the partition function $Z_\sigma = \exp(1+\lambda_0)$.

By a similar token, $S_{\mathcal{T}}[p(\bm{s}_t)]$, regarded as a higher-order functional, can be extremized with respect to $\pcalt(\bm{s}_t)$---the process that constitutes the \textit{principle} of maximum caliber. Constraints imposed on macroscopic variables $\bar{C}_{i=0,1,2} = \avg{C_i(\bm{s}_t)}$ with the Lagrange multipliers $\lambda_{i=0,1,2}$, also including the condition of normalized probabilities, yield~\cite{Monthus2011,Press2013,Ghosh2020}
\begin{equation}
    \pcalt(\bm{s}_t) \propto \exp\left(-\sum_{i=0,1,2,...} \lambda_i C_i(\bm{s}_t)\right).
\end{equation}
A variable $C_1(\bm{s}_t) = Q(\bm{s}_t)$ often proposed to describe statistics of nonequilibrium steady states~\cite{Filyukov1967I,Monthus2011}, is the time-averaged heat transfer $$
    Q(\bm{s}_t) = \int_\mathcal{T} \frac{dt}{\mathcal{T}}\, q(\bm{s}_t),
$$
in which $q(\bm{s}_t)$ is the instantaneous value of the heat current.

\subsection{\label{app:boltzmann}Boltzmann distribution}
Here we summarize the modern formulation of the microcanonical distribution for the simplified model of Boltzmann gas consisting of indistinguishable particles~(Sec.~\ref{sec:ideal}), e.g.~Ref.~\cite[Chapters 12 and 13]{Carter2001}. A distribution of particles over energy levels $i=1,2,...,M$ is specified by occupations numbers $\{n_i\}_{i=1}^M$ with the count of realizations $W$ given by the Maxwell-Boltzmann statistics
$$W \approx \prod_{i=1}^M \frac{g_i^{n_i}}{n_i!}.$$
Assuming that all realizations are equivalent, we extremize the objective function
$$
    f_W = \ln W + \alpha (N - \sum_i n_i) + \beta (E - \sum_i \epsilon_i n_i),
$$
where the Lagrange multipliers $\alpha$ and $\beta$ incorporate the constraints on the total number of particles $N$ and their energy $E$. By extremizing $f_W$, we get
\begin{equation}\label{eq:ni_eq}
    n_i = g_i e^{-\alpha - \beta \epsilon_i}.
\end{equation}
The Lagrange multiplier $\alpha$ can be evaluated in this model explicitly from the constraint
$$
    N = \sum n_i = e^{-\alpha} Z_{\rm MB},
$$
which yields $\alpha = - \ln (N / Z_{\rm B})$. Now we identify the Boltzmann entropy with
$$
    S_{\rm B} = \kB \ln W = \kB N + \kB \beta E + \kB \alpha N,
$$
which yields the Euler equation when multiplied by $T$:
$$
    T S_{\rm B} = N \kB T + T E - \mu_{\rm MB} N,
$$
with the chemical potential $\mu_{\rm MB} = -\kB T \alpha$. By replacing the Lagrange multiplier $\alpha = - \beta\mu_{\rm MB}$ in Eq.~\eqref{eq:ni_eq}, we get Eq.~\eqref{eq:micro}.

Note that to obtain the Boltzmann probability distribution we need to normalize Eq.~\eqref{eq:micro}, which amounts to division by $N$ and yields the fraction of particles residing in the energy level $\epsilon_i$ in equilibrium:
$$
    p_i = \frac{g_i}{N} e^{\beta (\mu_{\rm MB} - \epsilon_i)} = \frac{g_i}{Z_{\rm MB}} e^{-\beta \epsilon_i}.
$$

\subsection{\label{app:master}From Markov chains to master equation}
The right hand-side of the Markov chain Eq.~\eqref{eq:chain} can be expanded as
$$
    n_j(t+dt) = n_j(t) + dt \partial_t n_j(t) + O(dt^2),
$$
from which we obtain
\begin{multline}
    \partial_t n_j + O(dt) = \frac{\sum_i p_{ji} n_i - n_j}{dt} \\
        = \sum_i \frac{p_{ji} n_i - p_{ij} n_j}{dt} = \sum_{i \ne j} \frac{p_{ji} n_i - p_{ij} n_j}{dt},
\end{multline}
where we use Eq.~\eqref{eq:num} for $n_j(t)$ and omit its explicit time-dependence for brevity. With Eq.~\eqref{eq:rate}, the limit $dt\to0$ of the above equation yields a master equation of a Markov process $n_i(t)$:
\begin{equation}
    \partial_t n_i(t) = \sum_{j\ne i} \left[k_{ij} n_j(t) - k_{ji} n_i(t)\right].
\end{equation}

\subsection{\label{app:fd}Microcanonical caliber of Fermi gas}
To derive Eq.~\eqref{eq:fd} for a given set of elements $n_{ji}$, we focus on a single target level~$j$, while traversing the source levels in the increasing order of the index $i=1,2,...$ first. The number of ways to allocate $n_{j1}$ of particles transferred from the level $i = 1$ is given by the Fermi-Dirac statistic~\cite[Chapter 13]{Carter2001}:
$$
    \Omega_{j1} = \frac{g_{j}!}{(g_{j}- n_{j1})! n_{j1}!}.
$$
After that, there remain
$$
    \Omega_{j2} = \frac{(g_{j}-n_{j1})!}{(g_{j}- n_{j1} - n_{j2})! n_{j2}!}
$$
ways to allocate $n_{j2}$ particles transferred from the level $i=2$. By induction, choosing a particular sequence of the levels $i = 1,2,...,M$ we obtain
\begin{multline}\label{eq:fd_omegaj}
    \Omega_{j} =\frac{g_{j}!}{\cancel{(g_{j}-n_{j1})!} n_{j1}!}
    \times \frac{\cancel{(g_{j}-n_{j1})!}}{(\cancel{g_{j}-n_{j1}-n_{j2})!}n_{j2}!}
    \times \dots \\\times \frac{\cancel{(g_{j}-\sum_{i < M} n_{ji})!}}{(g_{j}-\sum_{i} n_{ji})! n_{jM}!} = \frac{g_{j}!}{(g_{j}-\sum_{i}n_{ji})!\prod_{i} n_{ji}!}.
\end{multline}

Remarkably, in the final expression of Eq.~\eqref{eq:fd_omegaj} the dummy index $i$ is eliminated from the denominator by the summation operator in the first factor $(g_j - \sum_i n_{ji})!$, and by the product operator in the second factor $\prod_i n_{ji}$. As commutative operators they \textit{do not depend} on the order of operands $n_{ji}$ and, hence, neither on the sequence, in which we traverse the source levels.

Therefore the $M!$ possible permutations of the index $i$, which specify the traversal order of the source levels, are indistinguishable and yield the same number of realizations $\Omega_j$. As each target level $j$ contributes towards the total count $\Omega(\bm{s}_{dt})$ of the whole path's realizations independently, Eq.~\eqref{eq:fd}
follows from
$$
    \Omega(\bm{s}_{dt}) = \prod_j \Omega_j.
$$

To maximize the caliber of a Fermi-Dirac gas we make use of the Stirling approximation
\begin{multline}
    \ln\Omega_{\rm FD} \approx \sum_{j} \Biggl\{
        g_{j}\ln g_{j} - g_j
        \Biggr.\\
        - \left(g_{j}-\sum_{i}n_{ji}\right)
        \left[\ln\left(g_{j}-\sum_{i} n_{ji}\right) - 1\right]
        \\\Biggl.
        - \sum_{i} \left(n_{ji}\ln n_{ji} - n_{ji}\right)
    \Biggr\}.
\end{multline}
to form the objective function
\begin{multline}
    f_{\rm FD} = \ln \Omega_{\rm FD} + \beta\left(E - \sum_{ij} \epsilon_j n_{ji}\right) \\ + \sum_i\theta_i \left(n_i - \sum_{ij} n_{ji} \right),
\end{multline}
with Lagrange multipliers $\beta$ and $\theta_i$. The extremum condition $\partial f_{\rm FD} / \partial n_{ji} = 0$ then yields
\begin{equation}\label{eq:fdnji}
    n_{ji} = (g_{j}-n_{j}') e^{-\beta \epsilon_{j} - \theta_{i}}.
\end{equation}
with $n_j' = n_j(t+dt)$. By imposing the constraint Eq.~\eqref{eq:num} on $n_i(t) = \sum_j n_{ji}$, we evaluate
\begin{equation}\label{eq:fdtheta}
    e^{-\theta_i} = \frac{n_{i}}{\sum_j \left(g_j - n_j'\right) e^{-\beta\epsilon_j}}
\end{equation}
Equations~\eqref{eq:fdnji} and \eqref{eq:fdtheta} combined together produce Eq.~\eqref{eq:fdmax}.

\subsection{\label{app:inter}Simplified model of pairwise interactions}
Simplified models of the classical statistical mechanics, like those considered in Sec.~\ref{sec:framework}, can also be generalized to analyze multibody interactions between particles and are not limited to hard-core repulsion. Here we discuss briefly how to introduce pairwise interaction into the Boltzmann model of ideal gas (Sec.~\ref{sec:ideal}). To this end we extend the structure of energy levels now numbered by two indices $a=1,2$ and $b=1,2,...,M$, where $a=1$ labels isolated particles and $a=2$ represents a doublet of interacting particles.

The occupation numbers $n_{1b}$ and $n_{2b}$ may be regarded as specifying two distinct types of particles, with the conversion between the two characterized by a chemical reaction $2 n_{1b} \leftrightarrow n_{2b'}$. The conservation of matter and energy takes then the following form of macroscopic constraints
\begin{equation}
    N = \sum_{ab} a n_{ab},\qquad E = \sum_{ab} a\epsilon_{ab} n_{ab},
\end{equation}
in which $2\epsilon_{2b}$ stands for the energy of a doublet, which is shared equally between the constituent particles. The transition $n_{ij|ab}$ from an energy level $\epsilon_{ab}$ to $\epsilon_{ij}$ now requires four indices. The constraint \eqref{eq:num} then should be recast as
$$
    a n_{ab}(t) = \sum_{ij} i n_{ij|ab},
    \quad
    i n_{ij}(t+dt) = \sum_{ab} a n_{ij|ab}.
$$

To take into account clusters of three or more interacting particles, energy levels $\epsilon_{ab}$ with $a=...3,4,...$ of triplets, quadruplets, etc. can be considered. Such a straightforward extension enables modeling of either pairwise-additive potentials, or more complex multibody interactions. This formalism resembles, in fact, the cluster expansion in the classical statistical mechanics~\cite{Mayer1941}, \cite[Sec. 5.2]{kardar2007statistical}.

\subsection{\label{app:diffusion}Coarse-graining and continuum limit of lattice models}
Here we derive a space-dependent diffusion equation for the system discussed in Sec.~\ref{sec:eq}, in order to illustrate two formal procedures applicable to lattice models, namely, coarse graining and taking the continuum limit of the sites' length scale $dx\to0$. In principle, the continuum limit can also be taken without coarse graining, leading  to reaction-diffusion equation, for brevity and simplicity of our example we combine both.

Although lattice models by themselves constitute a kind of coarse graining in space, in some physical contexts the information about distribution of particles over energy levels of the system is not accessible. In the problem of particles' diffusion, for example, this information is usually not taken into account. Using the spatial discretization we eliminate the distinction between energy levels at the lattice sites, by summing Eq.~\eqref{eq:chain0} describing a discrete random walk, over the index $j$:
\begin{equation}\label{eq:cg}
    \nu(m, t + dt) = \sum_{j} n_j(m, t + dt) = z(m) \sum_{\ell=m-1}^{m+1}\frac{\nu(\ell, t)}{\zeta(\ell)}.
\end{equation}
Applying the relation $$
    z(m) = \zeta(m) - z(m+1) - z(m-1)
$$ to Eq.~\eqref{eq:cg}, we get
\begin{multline}\label{eq:diffusion}
    \partial_t \nu(m, t) = - \kappa(m) \nu(m)\\+ k_{m,m+1}\nu(m+1) + k_{m,m-1}\nu(m-1),
\end{multline}
with transition rates $k_{m \pm 1,m} = z(m \pm 1) / \bm{(}\zeta(m) dt\bm{)}$ (\apx~\ref{app:master}), and escape rate $\kappa(m) = k_{m+1,m} + k_{m-1,m}$. Note that the transition rates satisfy a generalized detailed-balance relation
\begin{equation}\label{eq:rw}
    \frac{k_{m,\ell}}{k_{\ell,m}} = \frac{z(m) \zeta(m)}{z(\ell) \zeta(\ell)} = \frac{\zeta(m)}{\zeta(\ell)}e^{\mathcal{U}(m) - \mathcal{U}(\ell)},
\end{equation}
in which we identify a \textit{directing} function~\cite{Belousov2022}
$$
    \mathcal{U}(\ell) = \ln z(\ell).
$$
The directing function may emerge, for example, due to external fields, such as gravity.

Now, if we interpret lattice indices $\ell$ as spatial coordinates $x = \ell\,dx$, the random walk Eq.~\eqref{eq:rw} can be analyzed by standard methods~\cite[Chapters I and II]{Chandrasekhar1943}. In particular, due to the spatial dependency of $\kappa(x)$ the continuous limit of this random walk yields a Smoluchowski equation~\cite{Belousov2022} for the density $\rho(x) = \nu(x) / dx$:
\begin{equation}\label{eq:SE}
    \partial_t \rho(x) = \nabla \cdot \biggl[
        \beta D(x) \nabla U(x) \rho(x) + D(x) \nabla \rho(x)
    \biggr]
\end{equation}
with a space-dependent diffusion coefficient
$$
    \frac{\kappa(x) dx^2}{2} \underset{\substack{dx\to 0 \\ dt\to0}}{\to} D(x),
$$
and the effective potential
$$
    U(x) = \kB T \Bigl(\ln D(x) - 2 \mathcal{U}(x)\Bigr).
$$

In presence of nonequilbrium constraints, e.g. with a Lagrange multiplier $\eta \in \{\eta_{1L}, \eta_{\rm DM}, \eta_{\rm AD}\}$ (Secs.~\ref{sec:grad} and \ref{sec:act}), the coarse-grained Eq.~\eqref{eq:cg} takes a more general form:
\begin{equation}\label{eq:cgne}
    \nu_\ell(t + dt) = \sum_m p_{\ell m}(\eta) \nu_m(t),
\end{equation}
in which we simplify the notation $\nu_\ell(t) = \nu(\ell, t)$ and introduce the transition probabilities
\begin{equation}\label{eq:cgplm}
    p_{\ell m}(\eta) = \frac{z(m) w_{\ell m}(\eta)}{\zeta_\eta(m)},
\end{equation}
where the extended partition function $\zeta_\eta(m)$ and a weight $w_{\ell m}(\eta)$ may in general depend on the Lagrange multiplier $\eta$. For example, in the gradient-driven flow discussed in Sec.~\ref{sec:grad}, such a dependence appears in $p_{L1}$ and $p_{1L}$ with
\begin{align}
    w_{1L} = e^{-\eta_{1L}},\quad&
    \zeta_\eta(L) = z_{L - 1} + z_{L} + z_1 w_{1L},\\
    w_{L1} = e^{\eta_{1L}},\quad&
    \zeta_\eta(1) = w_{L1} z_{L} + z_{1} + z_2,
\end{align}
cf.~Sec.~\ref{sec:num}.

\subsection{\label{app:grad}Constraint of constant flux at the boundary}
By imposing the extremum condition $\partial f_{1L}/\partial n(mj|\ell i)$ on the objective function~\eqref{eq:fgrad} near the end points $m,\ell = \{1,L\}$, we get
\begin{align}\label{eq:end1}
    n(1j|Li) =& g_j(1) e^{-\beta\epsilon_j(1) - \theta_i(L) - \eta_{1L}},
    \\\label{eq:endL}
    n(Lj|1i) =& g_j(L) e^{-\beta\epsilon_j(L) - \theta_i(1) + \eta_{1L}},
\end{align}
whereas all the other transitions satisfy
\begin{equation}
    n(mj|\ell i) = g_j(m) e^{-\beta\epsilon_j(m) - \theta_i(\ell)}
\end{equation}
By evaluating the constraints associated with $\theta_i(\ell\not\in\{1,L\})$ we obtain Eq.~\eqref{eq:path0}. To evaluate $\theta_i(\ell\in\{1,L\})$ we apply Eq.~\eqref{eq:num0}:
\begin{align}
    n_i(1, t) =& \sum_j \Bigl(n(Lj|1i) + n(1j|1i) + n(2j|1i)\Bigr)\nonumber\\
        =& e^{-\theta_i(1)} \Bigl(e^{\eta_{1L}}z(L) + z(1) + z(2)\Bigr),\\
    n_i(L, t) =& \sum_j \Bigl(n(L-1,j|Li) + n(Lj|Li) + n(1j|Li)\Bigr)\nonumber\\
        =& e^{-\theta_i(L)} \Bigl(z(L-1) + z(L) + z(1)e^{-\eta_{1L}}\Bigr).
\end{align}
By solving the two above equations for the Lagrange multipliers $\theta_i(\ell\in\{1,L\})$, from Eqs.~\eqref{eq:end1} and \eqref{eq:endL} we obtain
\begin{align}\label{eq:n1jLi}
    n(1j|Li) =& \frac{g_j(1)}{\zeta_{\eta_{1L}}(L)} e^{-\beta\epsilon_j(1) + \phi_{1L}} n_{Li} = p(1j|Li,J) n_i(L),
    \\\label{eq:nLj1i}
    n(Lj|1i) =& \frac{g_j(L)}{\zeta_{\eta_{1L}}(1)} e^{-\beta\epsilon_j(L) + \phi_{L1}} n_{1i} = p(Lj|1i,J) n_j(L),
\end{align}
in which we introduce active exponents~\cite{Belousov2024} $\phi_{1L} = -\eta_{1L}$, $\phi_{L1} = \eta_{1L}$, and extended neighborhood partition functions
\begin{eqnarray}
    \zeta_{\phi}(1) &=&  e^{\phi_{L1}} z(L) + z(1) + z(2),\\
    \zeta_{\phi}(L) &=& z(L-1) + z(L) + z(1) e^{\phi_{1L}},
\end{eqnarray}
The detailed-balance condition for the rates $p(1j|Li,J)$ and $p(Lj|1i,J),\eta)$ is now broken by the active exponents. Note that the directing function~\cite{Belousov2022,Belousov2024} is $\mathcal{U}_j(\ell) = -\beta\epsilon_j(\ell) + \ln g_j$.

The constraint of constant flux implies an important assumption that either $0 < J < \nu(L)$ or $0 < - J < \nu(1)$. The value of the Lagrange multiplier $\eta_{1L}$ can be found once the structure of the energy levels is specified.

\subsection{\label{app:active}Active exponents}
Active exponents, which can be used to express the violation of the detailed-balance relations [Eq.~\eqref{eq:dbphi}] in nonequilibrium systems, in the continuum limit entail nonconservative forces and amplified fluctuations, e.g. as discussed in Ref.~\cite[Supplemental Material Sec. 2]{Belousov2024}. Here we relate this theory to another phenomenon---force-dependent diffusion. To do so, we consider a Th\'evenin ensemble of active Boltzmann particles, whose motion is enhanced by constant exponents
\begin{equation}
    \phi_{m\ell} = \begin{cases}
        \text{$\phi_+$ if $m=\ell+1$,}\\
        \text{$\phi_-$ if $m=\ell-1$,}\\
        \text{$0$ otherwise,}
    \end{cases}
\end{equation}
which can also be decomposed as
$$
    \phi_\pm = \Phi \pm \beta F dx / 2
$$
into the symmetric $\phi_+ + \phi_- = 2 \Phi$ and asymmetric $\phi_+ - \phi_- = \beta F dx$ parts, where $F dx$ is interpreted as the work done by the force $F$ acting on a particle along the path $dx$, cf.~Eqs.~\eqref{eq:dbphi} and \eqref{eq:thermo}. Then the master equation~\eqref{eq:diffusion} reads
$$
    \dot{\nu}_m = -\kappa \nu_m + k_- \nu_{m+1}  + k_+ \nu_{m-1},
$$
where $\nu_m(t)$ is the number of particles at the $m$\textsuperscript{th} lattice and overdot is the time derivative, whereas the rate constants are given by
$$
    k_\pm = \frac{1}{dt}\frac{e^{\phi_\pm}}{1 + e^{\phi_+} + e^{\phi_-}},\;
    \kappa = k_+ + k_- \frac{1}{dt}\frac{e^{\phi_+} + e^{\phi_-}}{1 + e^{\phi_+} + e^{\phi_-}}.
$$
By expanding $\nu_{m\pm1} = \nu_m \pm dx \nu_m' + dx^2 \nu_m'' / 2$ with prime standing for the spatial derivative, we get
\begin{equation}\label{eq:expand}
    \dot{\nu}_m = \frac{\kappa dx^2}{2} \nu_m'' - (k_+ - k_-) dx \nu_m',
\end{equation}
in which we used the identity $\kappa = k_+ + k_-$. Then we divide Eq.~\eqref{eq:expand} by $dx$ and substitute
\begin{align*}
    &\kappa = \frac{1}{dt}\, \frac{e^{\frac{\beta F dx}{2}} + e^{-\frac{\beta F dx}{2}}}{e^{-\Phi} + e^{\frac{\beta F dx}{2}} + e^{-\frac{\beta F dx}{2}}} \simeq
    \frac{2}{dt(2 + e^{-\Phi})},
    \\
    &k_+ - k_- = \frac{1}{dt}\, \frac{e^{\frac{\beta F dx}{2}} + e^{-\frac{\beta F dx}{2}}}{e^{-\Phi} + e^{\frac{\beta F dx}{2}} + e^{-\frac{\beta F dx}{2}}}
        \simeq \frac{\beta F dx}{dt(2 + e^{-\Phi})},
\end{align*}
where we neglect the terms that would lead to $O(dx^3)$, and take the continuum limit with
$$
    \lim_{dx\to0} \frac{dx^2}{dt (2 + e^{-\Phi})} = D(\Phi),\quad
    \lim_{dx\to0} \frac{\nu}{dx} = \rho,
$$
to obtain the Smoluchowski diffusion equation
\begin{equation}\label{eq:active_diffusion}
    \partial_t \rho = \nabla \cdot \big(D(\Phi) \nabla\rho - \beta D(\Phi) \bm{F}\big).
\end{equation}
The above equations put in evidence the nonconservative force $F$, and allow us to define an effective temperature $\tilde{\beta}$ by imposing the Einstein relation for the mobility $\beta D(\Phi) = \tilde{\beta} D(0)$, from which
\begin{equation}
    \tilde{\beta} = \frac{\beta D(\Phi)}{D(0)} \sim \frac{3 \beta}{2 + \exp(-\Phi)}
    .
\end{equation} For example, if we assume a simple force $\beta F =\lim_{dx\to 0} \hat\phi / dx$, which entails $\Phi = \hat\phi / 2$, we may transform Eq.~\eqref{eq:active_diffusion} into
\begin{multline}
    \partial_t \rho = \nabla\cdot \big(D(F) \nabla\rho - \beta D(F) \bm{F}\big)\\
    = \nabla\cdot \left(\frac{\tilde{\beta}(F)}{\beta} D(0) \nabla\rho - \tilde\beta(F) D(0) \bm{F}\right).
\end{multline}

\subsection{\label{app:algo} Simulating a constrained Markov chain}
Under equilibrium conditions, the stochastic process described by Eq.~\eqref{eq:markov} in discrete time can be simulated by using standard computational techniques. Each step of the simulation begins with a given distribution of particles $\bm{\nu}$. For each lattice site $\ell\in\{1,2,3\}$ we generate $\nu_\ell$ pseudorandom numbers $u_{i}(\ell)$ with $i=1,2,...,\nu_\ell$, which are uniformly distributed on the unit interval, i.e. $u_i(\ell) \in I = [0, 1)$. Iterating over each site $\ell$, we make partitions $I_{m}(\ell=1,2,3)$ of the interval $I$ into three \textit{bins} $m=1,2,3$ of size $p_{m\ell}$. Then, as if constructing a histogram, we determine the number of particles $n_{m\ell} = \#\bigl[u_i(\ell) \in I_{m}(\ell)\bigr]$ which are transferred from the site $\ell$ into the site $m$. The result of the simulation step is a new state $\bm{\nu'}$ with components $\nu_m' = \sum_{\ell=1}^3 n_{m\ell}$.

The described algorithm is also applicable in nonequilbirium simulations without explicit constraints, as in the case \ref{it:ne} of Sec.~\ref{sec:num}, since the matrix $\bm{p}$ remains invariant at all steps. However, to restrict the sampled numbers $n_{m\ell}$ in a certain way, we need to modify this algorithm.

First we discuss the simplest modification of the algorithm, which is necessary for simulating the directed motion of self-propelling particles (Sec.~\ref{sec:main})---case \ref{it:dir} considered in Sec.~\ref{sec:num}. Specifically, to ensure that in each step of the simulation $\sum_{\ell=1}^{3} n_{\ell+1,\ell} = J_{\rm DM}$, we adjust the boundaries of the bins $I_{m}(\ell)$ as follows. First we order these bins so that, $I_{\ell+1}(\ell)$ is the left-most subinterval in each partitioning of $I$, i.e.
$$\forall u \in I_{\ell+1}(\ell), u' \in I_{m\ne\ell+1}(\ell): u < u'.$$
Ordering of the other subintervals does not matter.

Then we pull all randomly generated numbers $u_i(\ell)$ together into a single sample
$$\bm{U} = \cup_{\ell=1, i=1}^{3,\nu_\ell} \{u_i(\ell)\}.$$
Because all elements of the form
\begin{equation}\label{eq:pp}
    p_{\ell+1,\ell}(\eta_{\rm DM}) = p_+(\eta_{\rm DM}) = (1 + 2 e^{\eta_{\rm DM}})^{-1},
\end{equation}
which determine the size of the bins $I_{\ell+1}(\ell)$, are equal, the number of elements $$\#[U_i \in \bm{U}: U_i < p_+(\eta_{\rm DM})]$$ counts the total number of particles jumping to the right from any site $\ell\in\{1,2,3\}$. So we may choose such $\eta_{\rm DM}$, that the interval $I_{+}(\eta_{\rm DM}) = [0, p_{+})$ of size $p_{+}$ contains precisely $J_{\rm DM}$ numbers $U_i \in \bm{U}$:
\begin{equation}\label{eq:trick}
\#[u \in \bm{U} \cap I_{+}(\eta_{\rm DM})] = J_{\rm DM}.
\end{equation}

In practice there is an interval of values $\eta_{\rm DM} \in \hat{I}$ that satisfy Eq.~\eqref{eq:trick}. This interval becomes narrower as the total number of particles increases $N\to\infty$. Therefore, if we sort numbers $U(i=1,2,...,N) \in \bm{U}$ in the increasing order, that is $U_i < U_{i+1}$, we may estimate
$$
    \eta_{\rm DM} = - \ln [2 p_+ / (1 - p_+)],
$$ with $$
    p_+ %= \frac{U(J_{\rm DM}) + U(J_{\rm DM} + 1)}{2}.
    = \bigl[U(J_{\rm DM}) + U(J_{\rm DM} + 1)\bigr] / 2.
$$

The described algorithm imposing a constant cumulative flux $J_{\rm DM}$ can be easily adapted to the simulations under the constraint of cumulative displacement $J_{\rm AD}$ (case \ref{it:act} in Sec.~\ref{sec:num}). Requiring $J_{\rm AD}$ particles out of the total $N$ to move is equivalent to fixing the number $J_0 = N - J_{\rm AD}$ to remain at the same site in each step. Therefore we reorder this time intervals $I_m(\ell)$, so that $I_\ell(\ell)$ is the left-most in the partitioning scheme. Noting that all the elements
\begin{equation}\label{eq:p0}
    p_{\ell\ell}(\eta_{\rm AD}) = p_0(\eta_{\rm AD}) = (1 + 2 e^{-\eta_{\rm AD}})^{-1}
\end{equation}
are equal, we obtain $$
    \eta_{\rm AD} = \ln [2 p_0 / (1 - p_0)]
$$
with
$$
    p_0 %= \frac{U(J_0) + U(J_0 + 1)}{2}.
        = \bigl[U(J_0) + U(J_0 + 1)\bigr] / 2.
$$

Modifications of the algorithm, which are required to simulate the system with a constant boundary flux (case~\ref{it:grad} in Sec.~\ref{sec:num}), are more complicated. We focus on the transition probabilities
\begin{equation}\label{eq:pnew}
    p_{1L} = (2 e^{\eta} + 1)^{-1},\qquad
    p_{L1} = (2 e^{-\eta} + 1)^{-1},
\end{equation}
which can be related by
$$
    p(\eta) = p_{1L}(\eta) = \frac{1 - p_{L1}(\eta)}{1 + 3 p_{L1}(\eta)}.
$$

Note that as the probability of particles jumping from the site $L$ to $1$ vanishes,
$p(\eta)\to 0$, all the particles at the site $1$ transit to $L$ in one simulation step. Thereby we observe the most negative possible flux $J_{\min} = -\nu_1$. As the probability $p(\eta) \to 1$, we observe the largest possible flux $J_{\max} = \nu_3$.

Now we merge the samples $u_i(1)$ and $u_i(L)$ from the first and third sites as
$$
    \bm{\Upsilon} = \cup_{i=1}^{\nu_1} \left\{\frac{1 - u_i(1)}{1 + 3 u_i(1)}\right\} \cup_{i=1}^{\nu_3} \{u_i(3)\},
$$
and order them in the increasing order $\Upsilon(i) < \Upsilon(i+1)$. Finally we choose $\eta = -\ln [2 p / (1 - p)]$ with
$$
    p = \frac{\Upsilon(J - J_{\min}) + \Upsilon(J - J_{\min} + 1)}{2}. 
$$

\bibliography{refs}% Produces the bibliography via BibTeX.

\end{document}